\documentclass[aps,preprint,amsmath,amssymb,showpacs,amsfonts]{revtex4}
\usepackage{epsfig}
\usepackage{graphicx}
\usepackage{dcolumn}
\usepackage{bm}

\allowdisplaybreaks[1]

\newcommand{\del}{\partial}

\newcommand{\ddsimple}[2]{\frac{\del #1}{\del #2}} 
\newcommand{\dd}[3]{\left(\frac{\del #1}{\del #2}\right)_{\! #3}} 

\newcommand{\Sym}{\operatorname{Sym}}
\newcommand{\Tr}{\operatorname{Tr}}

\begin{document}

\title{Anomalies in Superfluids and a Chiral Electric Effect}

\author{Yasha Neiman}
\email{yashula@gmail.com}
\author{Yaron Oz}
\email{yaronoz@post.tau.ac.il}
\affiliation{Raymond and Beverly Sackler School of Physics and Astronomy, Tel-Aviv University, Tel-Aviv 69978, Israel}

\date{\today}

\begin{abstract}
We analyze the chiral transport terms in relativistic superfluid hydrodynamics. In addition to the spontaneously broken symmetry current, we consider an arbitrary number of unbroken symmetries and extend the results of arXiv:1105.3733.  We suggest an interpretation of some of the new transport coefficients in terms of chiral and gravitational
anomalies. In particular, we show that with unbroken gauged charges in the system, one can observe a chiral electric conductivity - a current in a perpendicular direction to the applied electric field. We present a motivated proposal for the value of the associated transport coefficient, linking it to the triangle anomaly.
Along the way we present new arguments regarding the interpretation of the anomalous transport coefficients in normal fluids.
We propose a natural generalization of the chiral transport terms to the case of an arbitrary number of spontaneously broken symmetry currents.
\end{abstract}
\maketitle

\section{Introduction}

The most remarkable property of liquid helium below the $\lambda$-point is superfluidity. It is the ability of the fluid to flow inside narrow capillaries without friction, discovered by Kapitza \cite{Kapitza}. The theoretical basis for understanding the phenomenon of superfluidity was given by Landau \cite{Landau}. The hydrodynamics of a superfluid consists of two motions: the motion of the normal part of the fluid, and the motion of the superfluid part which is an irrotational one, i.e. its velocity is curl free \cite{Tisza,Landau}.
A superfluid can be described as a fluid with a spontaneously broken symmetry, where the superfluid component is the condensate, and its velocity is proportional to the Goldstone phase gradient. The hydrodynamics of relativistic superfluids has been studied in \cite{relativistic}, and is relevant to the study of neutron stars \cite{NStars} and highly dense quark matter at the low temperature Color-Flavor locked phase \cite{CFL} (for a general perspective on the CFL phase, see \cite{Alford:1997zt,Alford:2007xm}). A gravitational holographic dual description of relativistic superfluid hydrodynamics has been proposed in \cite{Bhattacharya:2011ee,Herzog:2011ec}.

Quantum anomalies in the microscopic gauge theory imply new transport terms in the fluid dynamics. The anomalous transport in normal (i.e. not super-) fluid dynamics has been studied in \cite{Son:2009tf,Neiman:2010zi,Loganayagam:2011mu,KharzeevSecondOrder:2011ds,Sadofyev:2010is}, and in the holographic gravitational framework in \cite{Erdmenger:2008rm,Banerjee:2008th,Eling:2010hu}. Experimental signatures of this anomalous transport were proposed in \cite{KerenZur:2010zw,Kharzeev:2010gr}.

The aim of this paper is to study chiral effects and anomalous transport in superfluid hydrodynamics. In a recent work \cite{Bhattacharya:2011tr}, the entropic constraints on superfluid transport terms were analyzed, and the allowed transport terms were listed (a partial list has been obtained also in \cite{Lin:2011mr}). This was done for the case of a single abelian spontaneously broken charge. In the present work, we extend the calculation of \cite{Bhattacharya:2011tr}, with a minor correction, to an arbitrary number of additional (possibly non-abelian) unbroken charges.

The observational importance of such an extension lies in the transport terms which involve gauge field strengths $F^a_{\mu\nu}$. For the broken charge, these are inserted as fictitious external fields, which serve to increase the power of the entropic argument. In reality, if the broken charge is gauged, i.e. in the superconducting case, the gauge fields will be dynamically excluded from the bulk of the system. However, if we have unbroken gauged charges alongside the broken charge, then their associated gauge fields may enter the superfluid, rendering these new transport terms observable. In particular, this comment applies to one of the transport terms, which may be called  a chiral electric effect:
\begin{align}
   J_{CEE}^{(1)a\mu} = c^a{}_{bc}\epsilon^{\mu\nu\rho\sigma}u_\nu\xi_\rho^b E^c_\sigma \ , \label{eq:chiralE}
\end{align}
compared to the standard electric conductivity term $J_{Conduct}^{a\mu} = \sigma^{ab}E_b^\mu$.

Here $c_{abc}$ are the transport coefficients, $u_\mu$ is the normal fluid's four-velocity, $\xi_\mu^a$ is the phase gradient of the broken symmetry, which is proportional to the velocity of the superfluid part, and $E^a_\mu$ is an electric field. We use Greek letters $(\mu,\nu,\dots)$ for spacetime indices, and Latin letters $(a,b,\dots)$ for charge indices. In \eqref{eq:chiralE} and later in eqs. \eqref{eq:T_1_short}-\eqref{eq:c}, we choose to present our results with a charge index over the phase gradient $\xi_\mu$. This is done in anticipation of an extension of the results to the case with multiple broken symmetries, and helps in clarifying the index structure of the transport coefficients. We stress, however, that our calculations strictly apply only to the case of a single $\xi_\mu$, and there is yet much to understand about phases with multiple broken charges, even in equilibrium.

We will present the results for the allowed transport coefficients differently from the authors of \cite{Bhattacharya:2011tr}. First, for concreteness, we will use the transverse fluid frame (for a thorough discussion of fluid frames, see \cite{Bhattacharya:2011ee}). Second, we will show that the results of \cite{Bhattacharya:2011tr} can be better organized by a different choice of variables, using $(s, n^a/s, \zeta^2)$ as a set of independent thermal parameters instead of the set $(s, \mu_a/T, \zeta^2/T^2)$ used in \cite{Bhattacharya:2011tr}. Here $s$ is the entropy density, $n^a$ the normal charge densities, $\mu_a$ the chemical potentials, $T$ the temperature and $\zeta$ the length of the component of $\xi_\mu$ transverse to $u^\mu$ \footnote{In the nonrelativistic case, $\zeta$ is proportional to the relative velocity of the normal and superfluid parts}. Third, we will group the transport coefficients in a way which is suggestive of their relation to anomalies.

Some general remarks are in order here. In conventional hydrodynamics, entropic considerations only serve to point out which transport terms are allowed. To find the actual form of the transport coefficients as functions of the state, one must resort to the microscopic theory. In general, transport coefficients are given by $n$-point functions in the microscopic theory, via relations known as Kubo formulas. In this respect, the calculation of \cite{Son:2009tf} was seminal: it derived an almost unique form for the chiral transport coefficients in a normal  fluid, in terms of the coefficient $C_{abc}$ of the $JJJ$ chiral triangle anomaly.

In \cite{Neiman:2010zi}, we noted that the entropic constraints on the normal-fluid chiral terms leave two arbitrary constants which were disregarded in \cite{Son:2009tf}. It was later noticed in \cite{Bhattacharya:2011tr} that one of these constants $\gamma$, multiplying a $T^3$ term in the anomalous current, is ruled out by CPT invariance. The other constant
$\beta_a$, multiplying a $T^2$ term in the anomalous current, was evaluated in \cite{Landsteiner:2011cp} using Kubo formulas for a theory of free fermions. It was found to be proportional to the coefficient of the $JTT$ gravitational triangle anomaly. This result was confirmed \cite{Landsteiner:2011iq} in a strongly-coupled holographic setup. It may then be suggested that $\beta_a$ is always related to the gravitational anomaly. Thus, certain transport coefficients may be in fact fixed by anomalies, even if the entropic constraints allow them to be more general.

In the normal fluid, the relationship between the chiral transport coefficients and the anomalous 3-point correlators is not entirely clear from the Kubo formula approach. There, the coefficients arise from 2-point correlators \cite{Amado:2011zx}. The 3-point correlator  arises effectively from the dependence of the fermion propagators on the chemical potentials $\mu_a$ or the temperature $T$. In the entropic approach of \cite{Son:2009tf}, the $JJJ$ anomaly comes in directly in the (non-)conservation law for $J_a^\mu$, but the final result for the transport coefficients is removed from this by a lengthy calculation. We will propose in section \ref{sec:normal} a new perspective on the normal-fluid chiral coefficients, which clarifies their relation to the triangle anomaly and to each other. This will be a preliminary step before discussing the superfluid case.

We will then extend this intuition to the new superfluid transport terms, after grouping them in a way which reveals the relevant structure. This leads us to propose a simplified form for the chiral constitutive relations at viscous order:
\begin{align}
   T_{chiral}^{(1)\mu\nu} ={}& \chi^a\pi^{(\mu}_\lambda\epsilon^{\nu)\lambda\rho\sigma}u_\rho\zeta_{a\sigma}
     + a^{abc}\zeta_a^{(\mu}\epsilon^{\nu)\rho\sigma\lambda}u_\rho\zeta_{b\sigma}\pi_{\lambda\kappa}\zeta_c^\kappa
     + b^{abc}_1\zeta_a^{(\mu}\epsilon^{\nu)\rho\sigma\lambda}u_\rho\zeta_{b\sigma}\hat E_{c\lambda} \label{eq:T_1_short} \\
 \begin{split}
   J_{chiral}^{(1)a\mu} ={}& \omega^\mu\left(C^{abc}\mu_b\mu_c + 2\beta^a T^2
      - \frac{2n^a}{h}\left(\frac{1}{3}C^{bcd}\mu_b\mu_c\mu_d + 2\beta^b\mu_b T^2\right) \right) \\
      &+ B_b^\mu\left(C^{abc}\mu_c - \frac{n^a}{h}\left(\frac{1}{2}C^{bcd}\mu_c\mu_d + \beta^b T^2 \right)\right) \\
      &+ b_2^{abc}\epsilon^{\mu\nu\rho\sigma}u_\nu\zeta_{b\rho}\pi_{\sigma\lambda}\zeta_c^\lambda + c^{abc}\epsilon^{\mu\nu\rho\sigma}u_\nu\zeta_{b\rho}\hat E_{c\sigma}
 \end{split} \label{eq:J_1_short} \\
 \nu_{chiral}^{(1)a} ={}& \frac{2}{h}\zeta^a_\mu\omega^\mu\left(\frac{1}{3}C^{bcd}\mu_b\mu_c\mu_d + 2\beta^b\mu_b T^2 \right)
    + \frac{1}{h}\zeta^a_\mu B_b^\mu\left(\frac{1}{2}C^{bcd}\mu_c\mu_d + \beta^b T^2\right) \label{eq:nu_1_short} \\
  s_{chiral}^{(1)\mu} ={}& {-\frac{\mu_a}{T}}J^{(1)a\mu}_{chiral} + \omega^\mu\left(\frac{1}{3T}C^{abc}\mu_a\mu_b\mu_c + 2\beta^a\mu_a T \right)
    + B_a^\mu\left(\frac{1}{2T}C^{abc}\mu_b\mu_c + \beta^a T \right)\ . \label{eq:s_1_short}
\end{align}
The transport terms are presented with the natural generalization to multiple broken charges, in order to highlight the index structure of the coefficients. However, we stress that terms unique to the case of multiple broken charges are not included. 

In \eqref{eq:T_1_short}-\eqref{eq:s_1_short}, $\zeta^a_\mu$ is the component of $\xi^a_\mu$ transverse to $u^\mu$; $h = \epsilon + p$ is the enthalpy density, where $p$ is the pressure and $\epsilon$ is the energy density; $\pi_{\mu\nu}$ is the shear tensor of $u_\mu$; $\omega^\mu = \frac{1}{2}\epsilon^{\mu\nu\rho\sigma}u_\nu \del_\rho u_\sigma$ is the axial vorticity, $B_a^\mu$ is the magnetic field, and $\hat E_a^\mu$ is the combination $\hat E_a^\mu \equiv E_a^\mu - TP^{\mu\nu}\nabla_\nu(\mu_a/T)$, where $P_\mu^\nu = \delta_\mu^\nu + u_\mu u^\nu$ is the projector orthogonal to $u^\mu$. $\nu^{(1)a}$ is the correction $u^\mu\xi^a_\mu - \mu^a$ to the Josephson equation. The transport coefficients $\chi^a$, $a^{abc}$, $b_1^{abc}$ and $b_2^{abc}$ are arbitrary functions of state, with $b_1^{abc} + b_2^{cba}$ satisfying an inequality with transport coefficients from the non-chiral sector. If the dynamics is time-reversal invariant, we have instead simply $b_1^{abc} + b_2^{cba} = 0$ due to the Onsager principle (see section \ref{sec:results:onsager}). The constant $C_{abc}$ is the coefficient of the chiral $JJJ$ anomaly, and the constant $\beta_a$ is (probably) the coefficient of the gravitational $JTT$ anomaly.

As for the chiral electric conductivity $c_{abc}$, we will argue that it's related to the $JJJ$ anomaly coefficient $C_{abc}$. In particular, we suggest that $c_{abc}$ in \eqref{eq:chiralE} and \eqref{eq:J_1_short} takes the form:
\begin{align}
 c^{abc} = C^{dbe}\left(\delta^a_d - \frac{n^a\mu_d}{h}\right)\left(\delta^c_e - \frac{n^c\mu_e}{h}\right) \ . \label{eq:c}
\end{align}

A few comments and comparison to \cite{Bhattacharya:2011tr}: The chiral electric conductivity $c_{abc}$ is denoted in \cite{Bhattacharya:2011tr} as $\tilde\kappa_{11}$. The coefficients $\tilde\eta$, $\tilde\kappa_{12}$, $\tilde\kappa_{21}$ and $\tilde\kappa_{22}$ from \cite{Bhattacharya:2011tr} (our $\chi^a$, $b_1^{abc}$, $b_2^{abc}$ and $a^{abc}$) seem to be unrelated to anomalies. The coefficient $\sigma_8$ (will be denoted $\alpha_{ab}$ in this paper) appears to be related to a $JJT$-type anomaly, which does not exist. We therefore expect this coefficient to vanish. The coefficient $\sigma_{10}$ from \cite{Bhattacharya:2011tr} (more precisely, $\sigma_{10} - 2(\mu/T)\sigma_8 - (C/2)(\mu/T)^2$) is a generalization of the $JTT$-type anomaly coefficient $\beta_0$, where the $0$ subscript denotes the broken charge. While for a normal fluid the entropic constraints set $\beta_0$ to a constant, for a superfluid they allow it to be an arbitrary function of state. The interpretation in terms of the $JTT$ anomaly suggests that the new freedom in the entropic constraints is spurious, and $\beta_0$ is in fact a constant. This conclusion and others are backed by several structural arguments, which are presented in section \ref{sec:interpret}.

The paper is organized as follows. Section \ref{sec:framework} defines our notations and the general framework of the calculation. In section \ref{sec:results}, we list the transport terms allowed by our calculation of the entropic constraints. The calculation is detailed in the Appendix.
In section \ref{sec:normal}, we present our interpretation of the known chiral transport terms in a normal fluid. In section \ref{sec:interpret}, we extrapolate from this our educated guesses regarding the interpretation and values of the new superfluid transport terms. Section \ref{sec:discuss} is devoted to a discussion and outlook.

\section{Framework} \label{sec:framework}

In this section, we outline the framework for the calculation of the chiral transport terms allowed by the entropic constraint 
for a superfluid with a single broken charge and arbitrary unbroken charges. The details of the calculation are given in the Appendix, while the results
are presented in section III.

\subsection{Definitions and equations of motion}

We consider a superfluid with arbitrary unbroken currents $J_i^\mu$ and a single spontaneously broken $U(1)$ current $J_0^\mu$. The full set of currents is collectively denoted as $J_a^\mu$. The structure constants of the charge algebra are $f_{abc}$ (out of which only $f_{ijk}$ may be nonzero). The thermal state at each point is determined by the normal velocity $u^\mu$, the temperature $T$, the phase gradient $\xi_\mu$ of the broken symmetry and the chemical potentials $\mu_i$ of the unbroken symmetries. We denote the timelike norm of $\xi_\mu$ by $\xi = \sqrt{-\xi_\mu\xi^\mu}$. In equilibrium, $\xi_\mu$ is related to the chemical potential $\mu_0$ by the Josephson condition $u^\mu\xi_\mu = \mu_0 + O(\varepsilon)$, where $\varepsilon$ is a formal small parameter whose powers indicate the number of gradients involved. We denote the transverse part of $\xi_\mu$ as :
\begin{align}
 \zeta_\mu \equiv P_\mu^\nu\xi_\nu = \xi_\mu + \mu_0 u_\mu;\quad \zeta^2 \equiv \zeta_\mu\zeta^\mu = \mu_0^2 - \xi^2 \ ,
\end{align}
where $P_\mu^\nu = \delta_\mu^\nu + u_\mu u^\nu$ is the projector orthogonal to $u^\mu$. The thermodynamic identities read:
\begin{align}
 dp &= s dT + n^a d\mu_a + \frac{1}{2}Q d\xi^2 \label{eq:first_law} \\
 h &= \epsilon + p = Ts + \mu_a n^a \ , \label{eq:enthalpy}
\end{align}
where $p$ is the pressure, $s$ is the entropy density, $\mu_a$ are the chemical potentials, $n_a$ are the normal charge densities, $\epsilon$ is the energy density, $h = \epsilon + p$ is the enthalpy density, and $Q$ is the (unnormalized) superfluid charge density. The ideal stress tensor, charge current and entropy current read:
\begin{align}
 T^{(0)\mu\nu} &= \epsilon u^\mu u^\nu + p P^{\mu\nu} + Q\xi^\mu\xi^\nu \\
 J_i^{(0)\mu} &= n_i u^\mu \\
 J_0^{(0)\mu} &= n_0 u^\mu - Q\xi^\mu \\
 s^{(0)\mu} &= su^\mu\ .
\end{align}
We take the metric $g_{\mu\nu}$ to be curved on the scale of the hydrodynamic gradients, with a Riemann tensor $R_{\mu\nu\rho\sigma} = O(\varepsilon^2)$. We couple an external gauge field $A^a_\mu$ to every current that is not already coupled to one. See \cite{Neiman:2010zi} for the subtleties involved in this procedure in the presence of anomalies; as explained there, we use the covariant version of the currents and a symmetric anomaly coefficient $C_{abc}$. For the broken current $J_0^\mu$, the introduction of the external field upgrades $\xi_\mu = -\del_\mu\phi + A^0_\mu$ from a phase gradient to an arbitrary covector. Its curl equals the corresponding field strength:
\begin{align}
 F^0_{\mu\nu} = 2\del_{[\mu}\xi_{\nu]} \ .
\end{align}
 We decompose the field strengths $F^a_{\mu\nu}$ into electric and magnetic pieces as:
\begin{align}
 F^a_{\mu\nu} = 2u_{[\mu} E^a_{\nu]} + B^a_{\mu\nu};\quad E^a_\mu = F^a_{\mu\nu}u^\nu;\quad B^a_{\mu\nu} = P_\mu^\rho P_\nu^\sigma F^a_{\rho\sigma}\ . \label{eq:F}
\end{align}
We also define the axial magnetic field vector:
\begin{align}
  B^{a\mu} = \frac{1}{2}\epsilon^{\mu\nu\rho\sigma}u_\nu F^a_{\rho\sigma} = \frac{1}{2}\epsilon^{\mu\nu\rho\sigma}u_\nu B^a_{\rho\sigma} \ .
\end{align}
We define a covariant derivative $\nabla_\mu$ which takes into account both the curved metric and the gauge fields. We decompose the velocity gradients $\nabla_\mu u_\nu$ into an acceleration $a^\mu$, a shear tensor $\pi_{\mu\nu}$, a vorticity tensor $\omega_{\mu\nu}$ and an expansion rate $\nabla_\mu u^\mu$:
\begin{align}
 \nabla_\mu u_\nu &= -u_\mu a_\nu + \pi_{\mu\nu} + \omega_{\mu\nu} + \frac{1}{3}\nabla_\rho u^\rho P_{\mu\nu} \label{eq:nabla_u} \\
 a^\mu &= u^\nu\nabla_\nu u^\mu \\
 \pi_{\mu\nu} &= P_\mu^\rho P_\nu^\sigma \nabla_{(\rho}u_{\sigma)} - \frac{1}{3}\nabla_\rho u^\rho P_{\mu\nu} \\
 \omega_{\mu\nu} &= P_\mu^\rho P_\nu^\sigma \nabla_{[\rho}u_{\sigma]}\ .
\end{align}
We also define the axial vorticity vector:
\begin{align}
 \omega^\mu = \frac{1}{2}\epsilon^{\mu\nu\rho\sigma}u_\nu \del_\rho u_\sigma = \frac{1}{2}\epsilon^{\mu\nu\rho\sigma}u_\nu \omega_{\rho\sigma} \ .
\end{align}
The ideal equations of motion read:
\begin{align}
 \nabla_\nu T_\mu^{(0)\nu} &= F^a_{\mu\nu} J_a^{(0)\nu} + O(\varepsilon^2) \label{eq:div_T_0} \\
 \nabla_\mu J_a^{(0)\mu} &= O(\varepsilon^2) \label{eq:div_J_0} \\
 u^\mu\xi_\mu &= \mu_0 + O(\varepsilon)\ . \label{eq:Josephson_0}
\end{align}
The ideal conservation laws \eqref{eq:div_T_0}-\eqref{eq:div_J_0} can be written as:
\begin{align}
 \nabla_\mu(n_0 u^\mu) &= \nabla_\mu\left(Q\xi^\mu \right) + O(\varepsilon^2) \label{eq:div_xi} \\
 \nabla_\mu(n_i u) &= O(\varepsilon^2) \label{eq:div_n} \\
 \nabla_\mu(su^\mu) &= O(\varepsilon^2) \label{eq:div_s} \\
 \begin{split}
   a^\mu &= \frac{1}{h}\left(n^a\hat E_a^\mu - \zeta^\mu\nabla_\nu(n_0 u^\nu)\right) - \frac{1}{T}P^{\mu\nu}\del_\nu T\ .
 \end{split} \label{eq:a}
\end{align}
To obtain eq. \eqref{eq:a} for the acceleration, we used the identities \eqref{eq:first_law}-\eqref{eq:enthalpy}. This is a slightly nonstandard expression, which has some advantages and simplifies our calculation in the Appendix. $\hat E_a^\mu$ is a combination of the electric field and the chemical potential gradient:
\begin{align}
 \hat E_a^\mu \equiv E_a^\mu - TP^{\mu\nu}\nabla_\nu\frac{\mu_a}{T} \ . \label{eq:E_hat}
\end{align}
This is the expression that arises in the standard normal fluid electric conductivity term $J_{Conduct}^{a\mu} = \sigma^{ab}\hat E_b^\mu$. As we will see in section \ref{sec:framework:second_law}, it plays a role in the entropy constraint. In section \ref{sec:interpret}, we will give it an interpretation in the thermal-QFT picture.

The viscous-order equations of motion read:
\begin{align}
 \nabla_\nu(T_\mu^{(0)\nu} + T_\mu^{(1)\nu}) &= F^a_{\mu\nu}(J_a^{(0)\nu} + J_a^{(1)\nu}) \label{eq:div_T} \\
 \nabla_\mu(J_a^{(0)\mu} + J_a^{(1)\mu}) &= C_{abc} E^b_\mu B^{c\mu} \label{eq:div_J} \\
 u^\mu\xi_\mu &= \mu_0 + \nu^{(1)}\ , \label{eq:Josephson}
\end{align}
where $T_\mu^{(1)\nu}$ is the first-order correction to the stress tensor, $J^{\mu(1)}$ is the first-order correction to the current, and $\nu^{(1)}$ is the first-order correction to the Josephson equation. $C_{abc} = C_{(abc)}$ is a constant tensor of anomaly coefficients. We work in the transverse frame, defined by:
\begin{align}
 u_\mu T^{(1)\mu}_\nu = 0; \quad u_\mu J_a^{(1)\mu} = 0  \ . \label{eq:Landau}
\end{align}

\subsection{Entropic constraints on the chiral terms} \label{sec:framework:second_law}

Introducing the correction $s^{(1)\mu}$ to the entropy current, the second law of thermodynamics reads:
\begin{align}
 \nabla_\mu(s^{(0)\mu} + s^{(1)\mu}) \geq 0  \ . \label{eq:second_law_raw}
\end{align}
We can use eqs. \eqref{eq:first_law}-\eqref{eq:enthalpy} and \eqref{eq:Josephson} to express $\nabla_\mu s^{(0)\mu}$ in terms of $\nabla_\nu T_\mu^{(0)\nu}$, $\nabla_\mu J_a^{(0)\mu}$ and $\nu^{(1)}$. We can then eliminate $\nabla_\nu T_\mu^{(0)\nu}$ and $\nabla_\mu J_a^{(0)\mu}$ using eqs. \eqref{eq:div_T}-\eqref{eq:div_J}. Eq. \eqref{eq:second_law_raw} is then written as a sum of manifestly second-order terms:
\begin{align}
 \begin{split}
   & \frac{1}{T}\left(-T^{(1)\mu\nu}\nabla_\mu u_\nu + J_a^{(1)\mu}\hat E^a_\mu + \nu^{(1)}\nabla_\mu(n_0 u^\mu) - C_{abc}\mu^a E^b_\mu B^{c\mu}\right) \\
   &{} + \nabla_\mu\left(s^{(1)\mu} + \frac{\mu^a}{T}J_a^{(1)\mu}\right) \geq 0  \ , \label{eq:second_law}
 \end{split}
\end{align}
or, equivalently:
\begin{align}
 \begin{split}
   & \frac{1}{T}\left(-T^{(1)\mu\nu}\pi_{\mu\nu} + \frac{1}{3s}T^{(1)\mu}_\mu u^\nu\del_\nu s + J_a^{(1)\mu}\hat E^a_\mu + s\nu^{(1)}u^\mu\del_\mu\frac{n_0}{s}
     - C_{abc}\mu^a E^b_\mu B^{c\mu}\right) \\
   &{} + \nabla_\mu\left(s^{(1)\mu} + \frac{\mu^a}{T}J_a^{(1)\mu}\right) \geq 0  \ . \label{eq:second_law2}
 \end{split}
\end{align}
We now wish to find the allowed terms in $T_\mu^{(1)\nu}$, $J_a^{(1)\mu}$, $\nu^{(1)}$ and $s^{(1)\mu}$ containing $\epsilon^{\mu\nu\rho\sigma}$. For normal fluids, the contribution of such chiral terms to the entropy production rate \eqref{eq:second_law} does not mix with the contribution from non-chiral terms. Furthermore, because its sign cannot be constrained, the contribution of the chiral terms to \eqref{eq:second_law} must vanish. The situation is not so simple in the superfluid case, as was noted in \cite{Bhattacharya:2011tr}. This is because chiral contributions to \eqref{eq:second_law} of the form $\epsilon^{\mu\nu\rho\sigma}U_\mu u_\nu\zeta_\rho V_\sigma$, where $U_\mu$ and $V_\mu$ are some first-order vectors, can mix with non-chiral contributions of the form $U_\mu U^\mu$, $U_\mu V^\mu$ and $V_\mu V^\mu$. More specifically, the non-chiral contributions may be positive semi-definite with a magnitude that is always greater or equal to the magnitude of $\epsilon^{\mu\nu\rho\sigma}U_\mu u_\nu\zeta_\rho V_\sigma$. The coefficient of the chiral contribution can then be nonvanishing, without violating the Second Law. The relevant vectors for the role of $U_\mu$ or $V_\mu$ are $\hat E^a_\mu$ (through the $J_a^{(1)\mu}\hat E^a_\mu$ term in \eqref{eq:second_law2}) and $\pi_{\mu\nu}\zeta^\nu$ (through the $T^{(1)\mu\nu}\pi_{\mu\nu}$ term in \eqref{eq:second_law2}).

\section{Results from the entropic constraints} \label{sec:results}

We derived the chiral transport terms allowed by the entropic constraint \eqref{eq:second_law2} for a superfluid with a single broken charge and arbitrary unbroken charges. The result reads:
\begin{align}
 \begin{split}
   T_{chiral}^{(1)\mu\nu} ={}& -sTP^{\mu\nu}\left(2T\zeta_\rho\omega^\rho\left(\frac{\mu_a}{T}\ddsimple{\alpha^a}{s} + \ddsimple{\beta_0}{s}\right)
     + \zeta_\rho B_a^\rho \ddsimple{\alpha^a}{s}\right) \\
    &- 2T\zeta^\mu\zeta^\nu\left(2T\zeta_\rho\omega^\rho\left(\frac{\mu_a}{T}\ddsimple{\alpha^a}{\zeta^2} + \ddsimple{\beta_0}{\zeta^2}\right)
     + \zeta_\rho B_a^\rho \ddsimple{\alpha^a}{\zeta^2} \right) \\
    &+ \chi\pi^{(\mu}_\lambda\epsilon^{\nu)\lambda\rho\sigma}u_\rho\zeta_\sigma + a\zeta^{(\mu}\epsilon^{\nu)\rho\sigma\lambda}u_\rho\zeta_\sigma\pi_{\lambda\kappa}\zeta^\kappa
     + b^a_1\zeta^{(\mu}\epsilon^{\nu)\rho\sigma\lambda}u_\rho\zeta_\sigma\hat E_{a\lambda}
 \end{split} \label{eq:T_1} \\
 \begin{split}
   J_{chiral}^{(1)a\mu} ={}& \omega^\mu\left(\vphantom{\frac{n^a}{h}} C^{abc}\mu_b\mu_c + 4\delta_0^{(a}\alpha^{b)}\mu_b T + 2\beta^a T^2 \right. \\
      &\left.{}\ - \frac{2n^a}{h}\left(\frac{1}{3}C^{bcd}\mu_b\mu_c\mu_d + 2\alpha^b\mu_b\mu_0 T + 2\beta^b\mu_b T^2 + \gamma T^3\right) \right) \\
      &+ B_b^\mu\left(C^{abc}\mu_c + 2T\delta_0^{(a}\alpha^{b)}
          - \frac{n^a}{h}\left(\frac{1}{2}C^{bcd}\mu_c\mu_d + 2\delta_0^{(b}\alpha^{c)}\mu_c T + \beta^b T^2 \right)\right) \\
      &+ 2T\zeta^\mu \left(\delta_0^a - \frac{\mu_0 n^a}{h} \right) 
       \left(2T\zeta_\nu\omega^\nu \left(\frac{\mu_b}{T}\ddsimple{\alpha^b}{\zeta^2} + \ddsimple{\beta_0}{\zeta^2}\right) 
       + \zeta_\nu B_b^\nu \ddsimple{\alpha^b}{\zeta^2} \right) \\
      &+ T\epsilon^{\mu\nu\rho\sigma}u_\nu\zeta_\rho\left(\frac{n^a T}{h}\del_\sigma\beta_0 - \left(\delta^a_b - \frac{n^a\mu_b}{h}\right)\nabla_\sigma\alpha^b\right) \\
      &+ b_2^a\epsilon^{\mu\nu\rho\sigma}u_\nu\zeta_\rho\pi_{\sigma\lambda}\zeta^\lambda + c^{ab}\epsilon^{\mu\nu\rho\sigma}u_\nu\zeta_\rho\hat E_{b\sigma}
 \end{split} \label{eq:J_1} \\
 \begin{split}
   \nu_{chiral}^{(1)} ={}& \zeta_\mu\omega^\mu\left(\frac{2}{h}\left(\frac{1}{3}C^{abc}\mu_a\mu_b\mu_c + 2\alpha^a\mu_a\mu_0 T + 2\beta^a\mu_a T^2 + \gamma T^3 \right) \right. \\
       &\left.{}\ + \frac{4T^2\mu_0\zeta^2}{h}\left(\frac{\mu_a}{T}\ddsimple{\alpha^a}{\zeta^2} + \ddsimple{\beta_0}{\zeta^2} \right)
        - \frac{2T^2}{s}\left(\frac{\mu_a}{T}\ddsimple{\alpha^a}{(n_0/s)} + \ddsimple{\beta_0}{(n_0/s)}\right) \right) \\
       {}+{}& \zeta_\mu B_a^\mu\left(\frac{1}{h}\left(\frac{1}{2}C^{abc}\mu_b\mu_c + 2\delta_0^{(a}\alpha^{b)}\mu_b T + \beta^a T^2\right)
        + \frac{2T\mu_0\zeta^2}{h}\ddsimple{\alpha^a}{\zeta^2} - \frac{T}{s}\ddsimple{\alpha^a}{(n_0/s)} \right)
 \end{split} \label{eq:nu_1} \\
 \begin{split}
   s_{chiral}^{(1)\mu} ={}& {-\frac{\mu_a}{T}}J^{(1)a\mu}_{chiral} + \omega^\mu\left(\frac{1}{3T}C^{abc}\mu_a\mu_b\mu_c + 2\alpha^a\mu_a\mu_0 + 2\beta^a\mu_a T + \gamma T^2 \right) \\
    &{} + B_a^\mu\left(\frac{1}{2T}C^{abc}\mu_b\mu_c + 2\delta_0^{(a}\alpha^{b)}\mu_b + \beta^a T \right)
        - T\epsilon^{\mu\nu\rho\sigma}u_\nu\zeta_\rho\left(\frac{\mu_a}{T}\nabla_\sigma\alpha^a + \del_\sigma\beta_0\right) \\
    &{} + \frac{\alpha_a}{2}\epsilon^{\mu\nu\rho\sigma}\xi_\nu F^a_{\rho\sigma}\ .
 \end{split} \label{eq:s_1}
\end{align}
Here $\chi$, $a$, $b_1^a$, $b_2^a$, $c_{ab}$, $\alpha_a$ and $\beta_0$ are arbitrary dimensionless functions of state, while $\beta_i$ and $\gamma$ are arbitrary dimensionless constants. The partial derivatives with respect to $s$, $n_0/s$ and $\zeta^2$ are taken with $(s, n_a/s, \zeta^2)$ as the independent thermal parameters; in other words, the derivative with respect to $n_0/s$ is taken at constant $s$, $n_i/s$ and $\zeta^2$, and so on. Note that a term of the form $\epsilon^{\mu\nu\rho\sigma}\del_\nu(x u_\rho\zeta_\sigma)$ can be added to \eqref{eq:s_1}, without changing the dynamics or the entropy production rate.

Our derivation of \eqref{eq:T_1}-\eqref{eq:s_1} is given in the Appendix. It is a generalized and streamlined version of the derivation in \cite{Bhattacharya:2011tr}. The combinations of transport coefficients which give a nonzero entropy production rate, and will have to be balanced by the non-chiral sector, are $b_1^a + b_2^a$ and $c_{[ab]}$. The former was noted in \cite{Bhattacharya:2011tr}, while the latter is specific to the case with multiple charges. However, we will see shortly that in a time-reversal-invariant theory, the Onsager principle imposes the relations:
\begin{align}
 b_2^a = -b_1^a;\quad c_{ab} = c_{ba} \label{eq:onsager}
\end{align}
This is precisely the condition for which the $b_1^a$, $b_2^a$ and $c_{ab}$ terms don't contribute to the entropy production rate. Thus, no counterbalancing from the non-chiral sector is required. Let us now demonstrate this relation.

\subsection{Onsager relations} \label{sec:results:onsager}

In \cite{Bhattacharya:2011tr}, it was argued that given time-reversal symmetry, the transport coefficients $b_1^a$ and $b_2^a$ (their $-\tilde\kappa_{12}$ and $\tilde\kappa_{21}$) should be related due to the Onsager principle \cite{onsager} as $\tilde\kappa_{21} = -\tilde\kappa_{12}$, or, in our terms, $b_2^a = b_1^a$. There appears to be a sign error in this relation, as we now show. 

We choose coordinates and a gauge so that locally $g_{\mu\nu} = \eta_{\mu\nu}$, $u^\mu = (1,0,0,0)$, $\zeta^\mu = (0,\zeta,0,0)$, and the Christoffel and gauge connection coefficients all vanish. Consider the charge $q_a$ and the momentum $p_x$ along the condensate's velocity $\zeta^\mu$. The conjugate quantities to $q_a$ and $p_x$ are the electric potential $\varphi_a$ and the velocity $v_x$, respectively. The gradient of $\varphi_a$ along the $y$-axis is $\del_y\varphi_a = -E_{ay}$, while the gradient of $v_x$ along the $z$-axis is $\del_z v_x = 2\pi_{zx}$, the latter equality holding when $\del_z v_x$ is the only nonzero velocity gradient. The current of $q_a$ along the $y$ axis is $J^a_y$, while the current of $p_x$ along the $z$ axis is $T_{xz}$. Leaving only the $b_1^a$ and $b_2^a$ terms in \eqref{eq:T_1}-\eqref{eq:J_1} (and setting the gradient $\del_y(\mu_a/T)$ to zero), we have:
\begin{align}
 T_{xz} &= \frac{\zeta^2}{2}b_1^a\epsilon^{ztxy}E_a^y = -\frac{\zeta^2}{2}b_1^a E_a^y = \frac{\zeta^2}{2}b_1^a\del_y\phi_a \\
 J^a_y &= \zeta^2 b_2^a\epsilon^{ytxz}\pi_{zx} = \zeta^2 b_2^a\pi_{zx} = \frac{\zeta^2}{2}b_2^a\del_z v_x
\end{align}
Thus, $(\zeta^2/2)b_1^a$ and $(\zeta^2/2)b_2^a$ are mirror-symmetric elements of the kinetic coefficient matrix. Now we must pay attention to the time-reversal properties of the relevant quantities. First, a time reversal flips the sign of $\zeta_x$, which is a property of the thermal state. However, in our context only the square of $\zeta_x$ enters, so this has no effect. Second, a time reversal flips the sign of $p_x$, but not of $q_a$. Therefore, the correct Onsager relation is an \emph{anti}symmetry of the kinetic coefficients, i.e. $b_2^a = -b_1^a$. A similar argument applied to the charge currents along the $y$ and $z$ axes shows that $c_{ab} = c_{ba}$. There, the sign-flip of $\zeta_x$ under time reversal cancels with the antisymmetry of $\epsilon^{\mu\nu\rho\sigma}$ with respect to the $y$ and $z$ axes.

\section{Revisiting the anomalous normal fluid} \label{sec:normal}

In this section, we present a new perspective on the known chiral transport terms for a normal fluid \cite{Son:2009tf,Neiman:2010zi}. For the arguments here and in section \ref{sec:interpret}, we regress to the abelian case. A careful non-abelian generalization is likely possible, as was done in \cite{Neiman:2010zi} for the arguments of \cite{Son:2009tf}.

Recall the transport terms in the charge current for a normal fluid \cite{Neiman:2010zi}:
\begin{align}
 \begin{split}
   J_a^{(1)\mu} ={}& \sigma_a^b\hat E_b^\mu + \omega^\mu\left(C^{abc}\mu_b\mu_c + 2\beta^a T^2
        - \frac{2n^a}{h}\left(\frac{1}{3}C^{bcd}\mu_b\mu_c\mu_d + 2\beta^b\mu_b T^2 + \gamma T^3\right)\right) \\
       &+ B_b^\mu\left(C^{abc}\mu_c - \frac{n^a}{h}\left(\frac{1}{2}C^{bcd}\mu_c\mu_d + \beta^b T^2 \right)\right) \ .
 \end{split} \label{eq:J_normal}
\end{align}
Here $C_{abc}$ is the coefficient of the $JJJ$ anomaly, $\beta_a$ is conjectured to be the coefficient of the $JTT$ anomaly, and $\gamma$ vanishes due to CPT invariance. In light of the progression of terms $C^{abc}$-$\beta^a$-$\gamma$, their charge index structure and the associated factors of $\mu_a$ and $T$, we can also associate $\gamma$ with the pure-gravitational $TTT$ anomaly. This interpretation again forces $\gamma$ to vanish, because such an anomaly doesn't exist in four-dimensional spacetime.

The combination $\hat E_a^\mu = E_a^\mu - TP^{\mu\nu}\nabla_\nu(\mu_a/T)$ in the electric conductivity term in \eqref{eq:J_normal} arises naturally in the context of the Second Law of thermodynamics. This comes about through the $\hat E^a_\mu J_a^{(1)\mu}$ term in the expression \eqref{eq:second_law} for the entropy production rate. The origin of the particular combination of vortical and magnetic terms in \eqref{eq:J_normal} is far less transparent. To improve this situation, we propose a certain heuristic way of looking at the transport terms. We note that the transport coefficients can be found, via Kubo formulas, from correlators in thermal QFT. These can be translated into Euclidean vacuum correlators, with a Euclidean metric $\tilde g_{\mu\nu}$, a $1/T$ periodicity in the imaginary time direction, and an external gauge potential $\mu^a u_\mu$. However, unlike in the Kubo approach, we will keep discussing non-equilibrium quantities such as $\omega^\mu$ and $B_a^\mu$ directly, instead of translating them into variations with respect to external fields.

With this approach in mind, we expect the system to respond not to the physical gauge potential $A^a_\mu$, but to the hybrid potential $\tilde A^a_\mu \equiv A^a_\mu + \mu^a u_\mu$. Consider the field strength $\tilde F^a_{\mu\nu} = 2\del_{[\mu}\tilde A^a_{\nu]}$ derived from this potential. Its electric and magnetic parts read:
\begin{align}
 \tilde E^a_\mu &\equiv \tilde F^a_{\mu\nu}u^\nu = E^a_\mu - P^\nu_\mu \del_\nu\mu^a - \mu^a a_\mu \label{eq:E_tilde} \\
 \tilde B_a^\mu &\equiv \frac{1}{2}\epsilon^{\mu\nu\rho\sigma}u_\nu\tilde F_{a\rho\sigma} = B_a^\mu + 2\mu_a\omega^\mu\ .
\end{align}
Using the ideal equation \eqref{eq:a} in the normal-fluid limit, we find that the electric part \eqref{eq:E_tilde} is in fact proportional to $\hat E^a_\mu$:
\begin{align}
 \tilde E^a_\mu = \left(\delta^a_b - \frac{\mu^a n_b}{h}\right)\hat E^b_\mu + O(\varepsilon^2) \ . \label{eq:E_relation}
\end{align}
Thus, the appearance of $\hat E^a_\mu$ in the electric conductivity term is consistent with the approach that $\tilde E^a_\mu$ is in fact the field to which the current reacts.

We should now look for the significance of $\tilde B_a^\mu = B_a^\mu + 2\mu_a\omega^\mu$ in the chiral terms of \eqref{eq:J_normal}. As a first step towards uncovering it, recall from the Kubo-formula analysis in \cite{Amado:2011zx} that the chiral terms in $J_a^\mu$ are the sum of a ``free'' piece and a piece multiplied by $n_a/h$:
\begin{align}
 J_{chiral}^{(1)a\mu} = J'^{a\mu} - \frac{n^a}{h}T'^\mu \ . \label{eq:Kubo}
\end{align}
The point of this decomposition is that the thermal-QFT correlators are directly related to $J'^\mu_a$ and $T'^\mu$, rather than to the full combination $J_{chiral}^{(1)a\mu}$. Furthermore, as their names are meant to suggest, $J'^\mu_a$ and $T'^\mu$ are related to the expectation values $\left<J_a^\mu\right>$ and $\left<T^{0\mu}\right>$ of the current and the stress tensor, respectively. The $T'^\mu$ term is associated with a reference-frame correction between the equilibrium and perturbed states of the fluid. Comparing \eqref{eq:Kubo} with \eqref{eq:J_normal}, we can decompose $J'^\mu_a$ and $T'^\mu$ into pieces proportional to $C_{abc}$, $\beta_a$ and $\gamma$:
\begin{align}
 \begin{split}
   J'^\mu_a &= C_{abc}J'^{bc\mu}_{(C)} + \beta_a J'^\mu_{(\beta)} \\
   T'^\mu &= C_{abc}T'^{abc\mu}_{(C)} + \beta_a T'^{a\mu}_{(\beta)} + \frac{\gamma}{2} T'^\mu_{(\gamma)} \ ,
 \end{split}
\end{align}
where the individual coefficients are given by:
\begin{align}
  & J'^{bc\mu}_{(C)} = \Sym\left\{\mu^b(B^{c\mu} + \mu^c\omega^\mu)\right\}; &
  & T'^{abc\mu}_{(C)} = \Sym\left\{\mu^a\mu^b\left(\frac{1}{2}B^{c\mu} + \frac{2}{3}\mu^c\omega^\mu\right)\right\}; \label{eq:J_C} \\
  & J'^\mu_{(\beta)} = 2T^2\omega^\mu; &
  & T'^{a\mu}_{(\beta)} = T^2(B^{a\mu} + 4\mu^a\omega^\mu); \label{eq:J_beta} \\
  & &
  & T'^\mu_{(\gamma)} = 4T^3\ . \label{eq:J_gamma}
\end{align}
Here, ``Sym'' denotes symmetrization over all charge indices. Consider now the factors \eqref{eq:J_C} associated with the $JJJ$ anomaly coefficient $C_{abc}$. We can write these as:
\begin{align}
  J'^{bc\mu}_{(C)} &= \Sym\int_0^\mu{d\mu^b(B^{c\mu} + 2\mu^c\omega^\mu)} = \int_0^\mu{d\mu^{(b}\tilde B^{c)\mu}}
    = \frac{1}{2}\epsilon^{\mu\nu\rho\sigma}\int_0^\mu{d(\mu^{(b} u_\nu)\tilde F^{c)}_{\rho\sigma}} \label{eq:J'_C} \\
  T'^{abc\mu}_{(C)} &= \Sym\int_0^\mu{\mu^a d\mu^b(B^{c\mu} + 2\mu^c\omega^\mu)} = \int_0^\mu{\mu^{(a} d\mu^b\tilde B^{c)\mu}}
    = \frac{1}{2}\epsilon^{\mu\nu\rho\sigma}\int_0^\mu{\mu^{(a} d(\mu^b u_\nu)\tilde F^{c)}_{\rho\sigma}}\ . \label{eq:T'_C}
\end{align}
The $u_\nu$ inside the integrals is understood to be constant; it is included with $\mu_b$ in the parentheses merely to emphasize the structure of the expression. We understand the contraction of \eqref{eq:J'_C} with $C_{abc}$ as a relation of the form $\delta J_a^\mu/\delta\tilde A^b_\nu \sim C_{abc}\epsilon^{\mu\nu\rho\sigma}\tilde F^c_{\rho\sigma}$, commonly encountered in the context of anomalies. In this interpretation, $\mu^a u_\mu$ acts as a constituent of the gauge potential $\tilde A^a_\mu$, as it should when the thermal state is translated into a Euclidean vacuum. The extra factor of $\mu_a$ in \eqref{eq:T'_C} as compared to \eqref{eq:J'_C} can also be understood to some extent. Introducing an external gauge potential affects the stress-energy operator by adding a ``potential energy'' term. In particular, the effective gauge potential $\mu^a u_\mu$ results in an addition $\delta T^{\mu\nu} = \mu^a\delta_0^{(\mu}J_a^{\nu)}$. Now, if for some reason it is only this addition that enters the chiral transport terms, then we have a heuristic explanation for the fact that \eqref{eq:T'_C} is the same as \eqref{eq:J'_C}, with $C^{abc}d\mu_b\tilde B_c^\mu$ replaced by $C^{abc}\mu_a d\mu_b\tilde B_c^\mu$. The integration in eqs. \eqref{eq:J'_C}-\eqref{eq:T'_C} can be interpreted as the gradual build-up of the relevant thermal-QFT correlators from the state with $\mu_a = 0$ to the state with $\mu_a \neq 0$. This point of view will be utilized in section \ref{sec:interpret:c}. 

Recently, the generalization of the $C_{abc}$ transport terms was obtained in arbitrary even spacetime dimensions: two papers \cite{Loganayagam:2011mu,KharzeevSecondOrder:2011ds} have found the generalization of $J'^{bc\mu}_{(C)}$, with \cite{KharzeevSecondOrder:2011ds} also giving the generalization of $T'^{abc\mu}_{(C)}$ (for the case of a single $U(1)$ charge). On inspection, the results of \cite{Loganayagam:2011mu,KharzeevSecondOrder:2011ds} satisfy suitably generalized versions of eqs. \eqref{eq:J'_C}-\eqref{eq:T'_C}. This lends credibility to our emphasis on these relations. For another recent clue regarding the role of $\tilde F^a_{\mu\nu}$ in the anomalous transport terms, see \cite{Sadofyev:2010is}.

The role of gravitational anomalies in the transport terms is less understood than that of the $JJJ$ anomalies. Here, we will make do with two modest observations. First, in the Euclidean picture, the temperature $T$ has a role with respect to metric variations similar to the role of $\mu_a$ with respect to gauge-potential variations. Indeed, we can consider Euclidean spacetimes with a fixed period in the time coordinate, but with different values of the metric component $\tilde g_{00}$. Since the inverse temperature $1/T$ corresponds to the metric length of the time period, we then have $\tilde g_{00} \sim 1/T^2$. This leads to the relation $d\ln\tilde g_{00} = -2d\ln T$, analogous to $d\tilde A^a_0 = d\mu^a$. Actually, if we wish to consider variations both in the metric and in the gauge potential, we should use $\mu_a/T$ rather than $\mu_a$ as the quantity associated with variations of $\tilde A^a_\mu$. This is because if we vary the metric $\tilde g_{\mu\nu}$ in a fixed coordinate system without varying $\tilde A^a_\mu$, what remains constant is not $\mu^a$, which is metric-normalized, but rather $\mu_a/T$, which is the chemical potential's contribution to the Aharonov-Bohm phase $\oint{\tilde A^a_\mu dx^\mu}$ over a period of the time coordinate. In section \ref{app:calc:cancellation} of the Appendix, we indeed see that $(T,\mu_a/T)$ is the cleanest choice of variables for deriving the anomalous terms in \eqref{eq:J_normal} and their superfluid generalization.

Our second observation is that the $\beta_a$ and $\gamma$ terms in \eqref{eq:J_normal} can be expressed in a form tightly related to our expressions \eqref{eq:J'_C}-\eqref{eq:T'_C} for the $C_{abc}$ terms. First, let us rewrite \eqref{eq:J'_C}-\eqref{eq:T'_C} in accord with the previous observation regarding the role of $\mu_a/T$ as opposed to $\mu_a$:
\begin{align}
  & \Sym\left\{T\tilde B^{b\mu}d\frac{\mu^a}{T}\right\} = dJ'^{ab\mu}_{(C)}; &
  & \Sym\left\{\mu^b T\tilde B^{c\mu}d\frac{\mu^a}{T}\right\} = dT'^{abc\mu}_{(C)}\ , \label{eq:J'_T'_C}
\end{align}
where $\mu_a/T$ is varied at constant $T$. Now we point out the following analogous relations for the $\beta_a$ and $\gamma$ terms in \eqref{eq:J_beta}-\eqref{eq:J_gamma}:
\begin{align}
  & J'^\mu_{(\beta)}d\frac{\mu^a}{T} = d(T\tilde B^{a\mu}); &
  & \Sym\left\{T'^{b\mu}_{(\beta)}d\frac{\mu^a}{T}\right\} = d(\mu^{(a} T\tilde B^{b)\mu}); \label{eq:J'_T'_beta} \\
  & &
  & T'^\mu_{(\gamma)}d\frac{\mu^a}{T} = dT'^{a\mu}_{(\beta)}\ . \label{eq:J'_T'_gamma}
\end{align}
This heuristic observation implies that the transport terms for $C_{abc}$, $\beta_a$ and $\gamma/2$ form a sequence, with a missing element between $C_{abc}$ and $\beta_a$. This missing element should be characterized by a constant rank-2 charge tensor $\alpha_{ab}$, and its transport terms should read:
\begin{align}
 & J'^{a\mu}_{(\alpha)} = T\tilde B^{a\mu}; &
 & T'^{ab\mu}_{(\alpha)} = \mu^{(a} T\tilde B^{b)\mu} \ . \label{eq:alpha_expectation}
\end{align}
We interpret this sequence as follows. $C_{abc} \sim \Tr\{G_{(a} G_b G_{c)}\}$ is the coefficient of the $JJJ$ anomaly, as we know explicitly from the entropic calculation; $\alpha_{ab} \sim \Tr\{G_{(a}G_{b)}\}$ is the would-be coefficient of the nonexisting $JJT$ anomaly (and is indeed absent from \eqref{eq:J_normal}); $\beta_a \sim \Tr\{G_a\}$ is the coefficient of the $JTT$ anomaly, as suggested in \cite{Landsteiner:2011cp} from a Kubo formula calculation; finally, $\gamma/2$ is the would-be coefficient of the nonexisting $TTT$ anomaly (which indeed must vanish in \eqref{eq:J_normal} due to CPT invariance). In the above, $G_a$ are the generators of the charge group in the fermions' representation.

A generalization of the $\beta_a$ and $\gamma$ contributions to $J'^\mu_a$ in arbitrary dimensions was recently discussed in \cite{Loganayagam:2011mu}, under the name of ``finite-temperature corrections'' to the anomalous transport terms. A sequence of such terms is found, its length dictated by the spacetime dimension. On inspection, this sequence is seen to obey a generalization of eqs. \eqref{eq:J'_T'_beta}-\eqref{eq:J'_T'_gamma}. Furthermore, the $C$-term in \cite{Loganayagam:2011mu} is related to the $(\beta,\gamma,\dots)$ sequence via a generalization of \eqref{eq:J'_T'_C}, again with a ``missing link'' between the $C$-term and the rest. We can again interpret the sequence in terms of polygon anomalies with a varying number of graviton vertices. The slot immediately after the $C$-term is always empty, because there is never an anomaly with a single graviton vertex.

\section{Interpretation and educated guesses for the superfluid transport terms} \label{sec:interpret}

In this section, we will use the insights from section \ref{sec:normal} to conjecture a more specific form for the transport terms \eqref{eq:T_1}-\eqref{eq:s_1}. This will bring us to the expressions \eqref{eq:T_1_short}-\eqref{eq:c} which were presented in the Introduction.

\subsection{Interpreting the $\alpha$ and $\beta$ terms} \label{sec:interpret:ugly}

Let us return to the superfluid result \eqref{eq:J_1} for the chiral part of the current $J_a^\mu$. We concentrate on the first two terms, involving the vorticity $\omega^\mu$ and the magnetic field $B_b^\mu$. We find that these terms reproduce the normal-fluid result \eqref{eq:J_normal}, with two differences. The first difference is that the coefficient $\beta_0$ is no longer a constant, but an arbitrary function of state. In the notation of \cite{Bhattacharya:2011tr}, it corresponds to $\sigma_{10} - 2(\mu/T)\sigma_8 - (C/2)(\mu/T)^2$. The second difference is the introduction of yet another arbitrary function of state, $\alpha_a$. In the notation of \cite{Bhattacharya:2011tr}, it corresponds to $\sigma_8$.

To understand better the role of $\alpha_a$, it will be useful to generalize to the case of multiple broken charges. We expect that the transport terms for a single broken charge will carry through to this more general case, with the trivial addition of a charge index on $\xi_\mu$. In addition, new transport terms are likely to appear, involving e.g. factors of $\epsilon^{\mu\nu\rho\sigma}u_\nu\zeta^a_\rho\zeta^b_\sigma$. In the present work, we disregard such new terms, and consider only the generalized versions of the terms \eqref{eq:T_1}-\eqref{eq:s_1}. In particular, we find that $\alpha_a$ gains a second charge index, becoming $\alpha_{ab}$; for instance, the last term in \eqref{eq:s_1} becomes $(\alpha_{ab}/2)\epsilon^{\mu\nu\rho\sigma}\xi^a_\nu F^b_{\rho\sigma}$. Following the derivation of the vortical and magnetic terms in the Appendix, we see that factors of $\alpha^{(ab)}$ come to replace factors of $\delta_0^{(a}\alpha^{b)}$. The $\alpha_{ab}$-contribution to the vortical and magnetic terms in the charge current reads:
\begin{align}
 J_{(\alpha)}^{a\mu} = 2\left(\alpha^{(ab)}T\tilde B_b^\mu - \frac{n^a}{h}\alpha^{(bc)}\mu_b\tilde B_c^\mu \right) \ . \label{eq:alpha_term}
\end{align}
Up to a factor of 2 which can be swallowed into the definition of $\alpha_{(ab)}$, this reproduces the expectation \eqref{eq:alpha_expectation} for the missing element in the sequence of coefficients between $C_{abc}$ and $\beta_a$.

It may seem at first that since $\alpha_a$ and $\beta_0$ in \eqref{eq:J_1} are arbitrary functions, the whole hierarchy of terms from section \ref{sec:normal} becomes meaningless. Indeed, why single out the arbitrary functions $\alpha_a$ and $\beta_0$? We might as well talk about all of $(1/3)C^{abc}\mu_a\mu_b\mu_c + 2\alpha^a\mu_a\mu_0 T + 2\beta^a\mu_a T^2 + \gamma T^3$ as an arbitrary function. However, $\alpha_a$ and $\beta_0$ really \emph{are} singled out by the entropic calculation. The evidence for this lies in other terms in the constitutive relations \eqref{eq:T_1}-\eqref{eq:nu_1}, which contain partial derivatives of $\alpha_a$ and $\beta_0$, rather than some other function, with respect to the thermal parameters $(s, n_0/s, \zeta^2)$. Furthermore, the derivatives of $\alpha_a$ and $\beta_0$ always come together, in the combination $\tilde B_a^\mu d\alpha^a + 2T\omega^\mu d\beta_0$. In the spirit of section \ref{sec:normal}, we note that the coefficients of this combination are related by $2T\omega^\mu d(\mu_a/T) = d\tilde B_a^\mu$. This reinforces the conclusion that $\alpha_a$ and $\beta_0$ (or, more generally, $\alpha_{(ab)}$ and $\beta_a$) indeed belong to the hierarchy of coefficients described in section \ref{sec:normal}, along with $C_{abc}$ and $\gamma$.

Now, from the entropic calculation we know that $C_{abc}$ and $\gamma$ remain constants in the superfluid case. Then the clean hierarchy of coefficients \eqref{eq:J'_T'_C}-\eqref{eq:alpha_expectation} suggests that $\alpha_{(ab)}$ and $\beta_a$ are constants as well. By this conjecture, all the partial derivatives of $\alpha_a$ and $\beta_0$ in \eqref{eq:T_1}-\eqref{eq:nu_1} vanish. For $\beta_0$, this implies a return to the normal-fluid situation: a constant $\beta_0$ is on a par with the other constant components of $\beta_a$, which are believed to be the coefficients of the $JTT$ anomaly. As for $\alpha_{(ab)}$, we can now interpret it as the would-be coefficient of the $JJT$ anomaly. Such an anomaly doesn't exist, which leads us to conjecture that in fact $\alpha_{(ab)} = 0$.

For the case of a single broken charge, we are done: $\alpha_{(ab)}$ collapses back to $\alpha_a$, which vanishes by the above argument. For multiple broken charges, we must also consider the antisymmetric piece $\alpha_{[ab]}$. Having come this far, it seems natural to guess that $\alpha_{[ab]}$ vanishes as well, so there are no partial-derivative terms in the constitutive relations at all. We note in this context that a \emph{constant} $\alpha_{[ab]}$ doesn't affect the dynamics; it only lends an identically non-dissipative term to the entropy current.

We end this subsection by summarizing the constitutive relations with our conjectures taken into account: the $\beta_a$ are all constants, and $\alpha_{ab}$ vanishes. We invoke CPT invariance to remove the $\gamma$ term, which has served its rhetorical role. Finally, the transport terms are presented with the naive generalization to multiple broken charges, in order to highlight the coefficients' index structure; we stress again that terms unique to the case of multiple broken charges are not included. The result is given in \eqref{eq:T_1_short}-\eqref{eq:s_1_short}.

\subsection{Interpreting the $c$ term: anomalous chiral electric effect} \label{sec:interpret:c}

Let us return to the representation \eqref{eq:J'_C}-\eqref{eq:T'_C} of the anomalous normal-fluid transport terms associated with $C_{abc}$. Let us consider these expressions in the context of a superfluid with a single broken charge. From the superfluid's point of view, the integration in \eqref{eq:J'_C}-\eqref{eq:T'_C} builds up from zero the longitudinal part $-\mu_0 u_\mu$ of $\xi_\mu$. The result gives us the anomalous currents in the presence of this longitudinal part. We propose that to obtain the full anomalous transport terms for the superfluid, one should proceed analogously to build up the transverse component $\zeta_\mu$ of $\xi_\mu$, at constant $\mu_a$. This will add the following contributions to $J'^{a\mu}$ and $T'^\mu$:
\begin{align}
 {J'}_{New}^{a\mu} &= -\frac{1}{2}P^\mu_\lambda C^{a0c}\epsilon^{\lambda\nu\rho\sigma}\int_0^{\zeta}{d\zeta_\nu\tilde F_{c\rho\sigma}}
   = -\frac{1}{2}P^\mu_\lambda C^{a0c}\epsilon^{\lambda\nu\rho\sigma}\zeta_\nu\tilde F_{c\rho\sigma} = C^{a0c}\epsilon^{\mu\nu\rho\sigma}u_\nu\zeta_\rho\tilde E_{c\sigma}
   \label{eq:J'_new} \\
 \begin{split}
   {T'}_{New}^\mu &= -\frac{1}{2}P^\mu_\lambda C^{a0c}\epsilon^{\lambda\nu\rho\sigma}\int_0^\zeta{\mu_a d\zeta_\nu\tilde F_{c\rho\sigma}}
      = -\frac{1}{2}P^\mu_\lambda C^{a0c}\mu_a\epsilon^{\lambda\nu\rho\sigma}\zeta_\nu\tilde F_{c\rho\sigma} \\
     &= C^{a0c}\mu_a\epsilon^{\mu\nu\rho\sigma}u_\nu\zeta_\rho\tilde E_{c\sigma}\ .
 \end{split} \label{eq:T'_new}
\end{align}
The projectors $P^\mu_\lambda$ are due to the transversality condition \eqref{eq:Landau}. The resulting contribution to the current reads:
\begin{align}
  J^{(1)a\mu}_{New} = {J'}_{New}^{a\mu} - \frac{n^a}{h}{T'}_{New}^\mu = C^{c0b}\left(\delta^a_c - \frac{n^a\mu_c}{h}\right)\epsilon^{\mu\nu\rho\sigma}u_\nu\zeta_\rho\tilde E_{b\sigma}\ .
  \label{eq:J_new_raw}
\end{align}
The relation \eqref{eq:E_relation} between $\tilde E^a_\mu = E^a_\mu - P^\nu_\mu\nabla_\nu\mu^a - \mu^a a_\mu$ and $\hat E^a_\mu = E^a_\mu - TP_\mu^\nu\nabla_\nu(\mu^a/T)$ doesn't quite hold in the superfluid case, because there is an extra $\zeta_\mu$-proportional term in the acceleration $a_\mu$. However, a weakened version of the relation does hold:
\begin{align}
 \tilde P_\mu^\nu\tilde E^a_\nu = \tilde P_\mu^\nu\left(\delta^a_b - \frac{\mu^a n_b}{h}\right)\hat E^b_\nu + O(\varepsilon^2)\ ,
\end{align}
where $\tilde P_\mu^\nu \equiv P_\mu^\nu - \zeta_\mu\zeta^\nu/\zeta^2$ is the projector onto the subspace orthogonal to both $u_\mu$ and $\xi_\mu$. We use this to rewrite eq. \eqref{eq:J_new_raw} as:
\begin{align}
  J^{(1)a\mu}_{New} = C^{c0d}\left(\delta^a_c - \frac{n^a\mu_c}{h}\right)\left(\delta^b_d - \frac{n^b\mu_d}{h}\right)\epsilon^{\mu\nu\rho\sigma}u_\nu\zeta_\rho\hat E_{b\sigma}\ . \label{eq:J_guess_0}
\end{align}
This has precisely the form of the $c_{ab}$-term from \eqref{eq:J_1}, with:
\begin{align}
 c_{ab} = C^{c0d}\left(\delta^a_c - \frac{n^a\mu_c}{h}\right)\left(\delta^b_d - \frac{n^b\mu_d}{h}\right)\ . \label{eq:c_guess}
\end{align}
We propose this as an educated guess for the chiral electric conductivity $c_{ab}$. It implies that the chiral electric transport term is directly related to the $JJJ$ anomaly. Note that our expression \eqref{eq:c_guess} satisfies the symmetry $c_{ab} = c_{ba}$, as required by the Onsager relation \eqref{eq:onsager}. 

The integration recipe in \eqref{eq:J'_C}-\eqref{eq:T'_C} and \eqref{eq:J'_new}-\eqref{eq:T'_new} doesn't seem to give the correct results if we first build up $\zeta_\mu$, and then $\mu_a$. This makes physical sense: states with $\mu_a = 0$ and nonzero $\zeta_\mu$ are forbidden, since $\xi_\mu$ must always be timelike in order to describe the superfluid velocity.

Generalizing our guess \eqref{eq:J_guess_0} to the case with non-abelian charges and multiple broken generators, we get an expression for the superfluid-specific addition to the anomalous transport terms:
\begin{align}
 J^{(1)a\mu}_{New} = C^{dbe}\left(\delta^a_d - \frac{n^a\mu_d}{h}\right)\left(\delta^c_e - \frac{n^c\mu_e}{h}\right)\epsilon^{\mu\nu\rho\sigma}u_\nu\zeta_{b\rho}\hat E_{c\sigma} \label{eq:J_guess} \ ,
\end{align}
where the index $b$ runs over the broken generators. This is equivalent to expression \eqref{eq:c} for the generalized chiral conductivity $c_{abc}$.

\subsection{Other terms}

The transport coefficients $\chi^a$, $a^{abc}$, $b_1^{abc}$ and $b_2^{abc}$ appear to be unrelated to anomalies. We expect anomalous terms to be associated with curvatures of the gauge and metric fields. This should make them proportional to antisymmetrized derivatives such as $\tilde F^a_{\mu\nu}$ and, for gravitational anomalies, $\omega_{\mu\nu}$. On the other hand, the terms corresponding to $(\chi^a, a^{abc}, b_1^{abc}, b_2^{abc})$ are associated with the \emph{symmetrized} derivative $\pi_{\mu\nu}$. This is not immediately clear for $b_1^{abc}$, but see \cite{Bhattacharya:2011tr} or our derivation in the Appendix for its close relationship with $b_2^{abc}$.

\section{Discussion} \label{sec:discuss}

We analyzed the chiral transport terms in relativistic superfluid hydrodynamics and extended the calculation of \cite{Bhattacharya:2011tr} to an arbitrary number of additional (possibly non-abelian) unbroken charges. We proposed an interpretation of some of the new transport coefficients in terms of chiral and gravitational anomalies. We showed that with unbroken gauged charges in the system, one can observe a chiral electric conductivity - a current in a perpendicular direction to the applied electric field. We proposed an explicit dependence of this conductivity on the anomaly coefficient $C_{abc}$. Finally, we presented a natural generalization of the chiral transport terms to the case of an arbitrary number of spontaneously broken symmetry currents.

There are several open issues for future work. Clearly, our conjectured simplification \eqref{eq:T_1_short}-\eqref{eq:s_1_short} of the transport terms and our proposal \eqref{eq:c} for the chiral electric conductivity should be tested with a microscopic calculation. Such a calculation with just one (broken) charge will already be a useful check. Also, it will be interesting to have explicit calculations, either thermodynamical or microscopic, for the transport terms in a superfluid with several broken charges. We expect this more general case to be relevant for nuclear and subnuclear fluids, where there are multiple potentially broken generators for the color and flavor symmetries.

The observational relevance of our results, and indeed of previous results along these lines, should be considered. As pointed out in \cite{Bhattacharya:2011tr}, the transport terms which aren't related to anomalies may have manifestations in nonrelativistic condensed-matter systems. Perhaps there is such hope for the anomalous terms as well - though the anomaly is a relativistic effect, so is magnetism; nonrelativistic velocities do not necessarily preclude the observation of such phenomena. On the particle-physics front, the currently known superfluid phases include neutron-star matter and the Color-Flavor locked phase of QCD. The CFL phase is particularly interesting for our purposes, since it includes an unbroken gauged generator, with respect to which the system behaves as an insulator \cite{Alford:2007xm}. This may offer an ideal setting for the chiral electric conductivity to be expressed: on one hand, there is an unbroken gauge field in the presence of other broken symmetries, as required for the effect; on the other hand, since the conventional conductivity vanishes, the chiral conductivity will have a strong signature.

\section*{Acknowledgements}		

We thank Shira Chapman for pointing out a minor error in the previous version. The work is supported in part by the Israeli Science Foundation center of excellence, by the US-Israel Binational Science Foundation (BSF), and by the German-Israeli Foundation (GIF).

\appendix

\section{Derivation of the transport terms from entropic constraints} \label{app:calc}

\subsection{The first-order terms allowed by symmetries and the ideal equations} \label{app:calc:terms}

Let us list all the algebraically distinct chiral terms that can appear in the constitutive relations:
\begin{align}
 \begin{split}
   T_{chiral}^{(1)\mu\nu} ={}& t_1 P^{\mu\nu}\zeta_\rho\omega^\rho + t^a_2 P^{\mu\nu}\zeta_\rho B_a^\rho + t_3 \zeta^\mu\zeta^\nu\zeta_\rho\omega^\rho
     + t^a_4 \zeta^\mu\zeta^\nu\zeta_\rho B_a^\rho + t_5 \zeta^{(\mu}\omega^{\nu)} + t^a_6 \zeta^{(\mu}B_a^{\nu)} \\
     &+ t_7 \zeta^{(\mu}\epsilon^{\nu)\rho\sigma\lambda}u_\rho\zeta_\sigma\del_\lambda T
     + t^a_8 \zeta^{(\mu}\epsilon^{\nu)\rho\sigma\lambda}u_\rho\zeta_\sigma\nabla_\lambda\frac{\mu_a}{T}
     + t_9 \zeta^{(\mu}\epsilon^{\nu)\rho\sigma\lambda}u_\rho\zeta_\sigma\del_\lambda\zeta^2 \\
     &+ t^a_{10} \zeta^{(\mu}\epsilon^{\nu)\rho\sigma\lambda}u_\rho\zeta_\sigma\hat E_{a\lambda}
     + t_{11} \zeta^{(\mu}\epsilon^{\nu)\rho\sigma\lambda}u_\rho\zeta_\sigma\pi_{\lambda\kappa}\zeta^\kappa
     + t_{12} \pi^{(\mu}_\lambda\epsilon^{\nu)\lambda\rho\sigma}u_\rho\zeta_\sigma \\
     &+ t_{13}g_{\kappa\lambda}\nabla^{(\mu}\xi^{\kappa}\epsilon^{\nu)\lambda\rho\sigma}u_\rho\zeta_\sigma
 \end{split} \label{eq:T_1_raw} \\
 \begin{split}
   J_{chiral}^{(1)a\mu} ={}& j^a_1 \omega^\mu + j^{ab}_2 B_b^\mu + j^a_3 \zeta^\mu\zeta_\nu\omega^\nu + j^{ab}_4 \zeta^\mu\zeta_\nu B_b^\nu
     + j^a_5 \epsilon^{\mu\nu\rho\sigma}u_\nu\zeta_\rho\del_\sigma T \\
     &+ j^{ab}_6 \epsilon^{\mu\nu\rho\sigma}u_\nu\zeta_\rho\nabla_\sigma\frac{\mu_b}{T} + j^a_7 \epsilon^{\mu\nu\rho\sigma}u_\nu\zeta_\rho\del_\sigma\zeta^2
     + j^{ab}_8 \epsilon^{\mu\nu\rho\sigma}u_\nu\zeta_\rho\hat E_{b\sigma} \\
     &+ j^a_9 \epsilon^{\mu\nu\rho\sigma}u_\nu\zeta_\rho\pi_{\sigma\lambda}\zeta^\lambda
 \end{split} \label{eq:J_1_raw} \\
 \nu_{chiral}^{(1)} ={}& \nu_1 \zeta_\mu\omega^\mu + \nu^a_2 \zeta_\mu B_a^\mu \label{eq:nu_1_raw} \\
 \begin{split}
   s_{chiral}^{(1)\mu} ={}& {-\frac{\mu_a}{T}}J^{(1)a\mu}_{chiral} + s_1 \omega^\mu + s^a_2 B_a^\mu + \frac{s^a_3}{2}\epsilon^{\mu\nu\rho\sigma}\xi_\nu F_{a\rho\sigma} \\
     &+ s_4 \epsilon^{\mu\nu\rho\sigma}u_\nu\xi_\rho\del_\sigma T + s^a_5 \epsilon^{\mu\nu\rho\sigma}u_\nu\xi_\rho\nabla_\sigma\frac{\mu_a}{T}
     + s_6 \epsilon^{\mu\nu\rho\sigma}u_\nu\xi_\rho\del_\sigma\zeta^2 \\
     &+ s_7 \xi^\mu\xi_\nu\omega^\nu + s^a_8 \xi^\mu\xi_\nu B_a^\nu + s^a_9 \epsilon^{\mu\nu\rho\sigma}u_\nu\xi_\rho E_{a\sigma}
     + s_{10} \epsilon^{\mu\nu\rho\sigma}u_\nu\xi_\rho\xi^\lambda\nabla_\sigma u_\lambda\ .
 \end{split} \label{eq:s_1_raw}
\end{align}
The functions $t_n$, $j_n$, $\nu_n$ and $s_n$ are candidate transport coefficients. For the argument that the set of terms \eqref{eq:T_1_raw}-\eqref{eq:s_1_raw} is complete and independent, we refer to \cite{Bhattacharya:2011tr}. The introduction of unbroken charges does not change the reasoning, beyond the trivial addition of charge indices. We've omitted from $s^{(1)\mu}$ a possible divergence-free term of the form $\epsilon^{\mu\nu\rho\sigma}\del_\nu(x u_\rho\zeta_\sigma)$, which does not affect the entropic constraints.

The candidate terms for $J_a^{(1)\mu}$ and $s^{(1)\mu}$ are written differently for two reasons. First, $J_a^{(1)\mu}$ obeys the transversality constraint \eqref{eq:Landau}, while $s^{(1)\mu}$ does not. Second, the form of the $J_a^{(1)\mu}$ terms was chosen for algebraic convenience, while the $s^{(1)\mu}$ terms were chosen so as to ease the calculation of their divergence.

\subsection{Contributions to the entropy production rate} \label{app:calc:rate}

We will now write the contributions to the entropy production rate \eqref{eq:second_law2} arising from \eqref{eq:T_1_raw}-\eqref{eq:s_1_raw}. To avoid equivalent terms written in two different ways, we will follow the following rules:
\begin{enumerate}
 \item We express $a^\mu$ in terms of other vectors, using the ideal equation \eqref{eq:a}.
 \item We express factors of $\nabla_\mu u^\mu$ and $\nabla_\mu\xi^\mu$ in terms of other scalars, using the ideal equations \eqref{eq:div_xi} and \eqref{eq:div_s}.
 \item We express the curl $2\del_{[\mu}\xi_{\nu]}$ as $F^0_{\mu\nu}$.
 \item We decompose factors of $\nabla_\mu u_\nu$ and $F^a_{\mu\nu}$ using eqs. \eqref{eq:nabla_u} and \eqref{eq:F}.
 \item We avoid the symmetrized derivative $\nabla_{(\mu}\xi_{\nu)}$ whenever possible, using the relations:
   \begin{align*}
     u^\nu\nabla_\mu\xi_\nu = \del_\mu\mu_0 - \xi^\nu\nabla_\mu u_\nu + O(\varepsilon^2);\quad \xi^\nu\nabla_\mu\xi_\nu = -\frac{1}{2}\del_\mu\xi^2
   \end{align*}
 \item We eliminate factors of $\xi^\mu\xi^\nu\pi_{\mu\nu}$, using the relation:
   \begin{align*}
     \xi^\mu\xi^\nu\nabla_\mu u_\nu = \xi^\mu\nabla_\mu\mu_0 - \xi^\mu u^\nu\nabla_\mu\xi_\nu + O(\varepsilon^2)
       = \xi^\mu\nabla_\mu\mu_0 - \xi^\mu E^0_\mu + \frac{1}{2}u^\nu\nabla_\nu\xi^2 + O(\varepsilon^2)
   \end{align*}
 \item We make sure that one index of each $\epsilon^{\mu\nu\rho\sigma}$ is always contracted with $u_\mu$, using identities of the form:
   \begin{align*}
    \begin{split}
      \epsilon^{\mu\nu\rho\sigma}x_\mu y_\nu z_\rho w_\sigma ={}& -x_\lambda u^\lambda \epsilon^{\mu\nu\rho\sigma}u_\mu y_\nu z_\rho w_\sigma
        - y_\lambda u^\lambda \epsilon^{\mu\nu\rho\sigma}x_\mu u_\nu z_\rho w_\sigma \\
        &- z_\lambda u^\lambda \epsilon^{\mu\nu\rho\sigma}x_\mu y_\nu u_\rho w_\sigma
        - w_\lambda u^\lambda \epsilon^{\mu\nu\rho\sigma}x_\mu y_\nu z_\rho u_\sigma
    \end{split}
   \end{align*}
  We don't carry out the analogous procedure with $\xi_\mu$, to maintain the standard form of the vorticity $\omega^\mu$ and the magnetic field $B_a^\mu$. This leads to a few instances of redundant forms for equivalent terms; these instances, however, are restricted to the $s_7$-$s_{10}$ contributions, which will be disqualified independently due to second-derivative terms.
  \item We bring all second derivatives to one of the distinct forms $\xi^\rho\nabla_\rho\nabla_\mu u_\nu$, $u^\rho\nabla_\rho F^a_{\mu\nu}$ and $\xi^\rho\nabla_\rho F^a_{\mu\nu}$.
  \item We note that the gradient of any thermal function can be written as a linear combination of some $k+2$ basic gradients, where $k$ is the number of charges. However, we will not bother to do so at this point. Two sets of independent gradients will be convenient at different stages of the calculation.
  \item We note that gradients of the form $u^\mu\nabla_\mu x$ are not independent, since we have $u^\mu\nabla_\mu(n_i/s) = 0$ from the ideal equations. At this point, we will leave them as they are.
\end{enumerate}
When commuting derivatives, we take into account the Riemann tensor $R_{\mu\nu\rho\sigma}$ and the gauge field strength $F^a_{\mu\nu}$ where necessary. In practice, the gauge field strength never comes up in this context. The only place where we must commute gauge-covariant derivatives is in the $s^a_5$ term, where we use the fact that $f_{abc}\mu^a s_5^b = 0$, since $s^a_5$ is a group-covariant function of $\mu_a$ plus scalars. The contributions to the entropy production rate read:
\begin{align*}
   & (\nabla_\mu s^\mu)_{chiral} = -C_{abc}\frac{\mu^a}{T}E^b_\mu B^{c\mu} \\
   &{} + \frac{t_1}{sT}\zeta_\mu\omega^\mu u^\nu\del_\nu s + \frac{t^a_2}{sT}\zeta_\mu B_a^\mu u^\nu\del_\nu s \\
   &{} + \frac{t_3}{T}\zeta_\mu\omega^\mu\left(\zeta^\nu\left(\hat E^0_\nu - \frac{\mu_0 n_a}{h}\hat E^a_\nu\right)
       + u^\nu\left(\frac{1}{2}\del_\nu\zeta^2 + \frac{\mu_0 s\zeta^2}{h}\del_\nu\frac{n_0}{s} \right)\right) \\
   &{} + \frac{t^a_4}{T}\zeta_\mu B_a^\mu\left(\zeta^\nu\left(\hat E^0_\nu - \frac{\mu_0 n_b}{h}\hat E^b_\nu\right)
       + u^\nu\left(\frac{1}{2}\del_\nu\zeta^2 + \frac{\mu_0 s\zeta^2}{h}\del_\nu\frac{n_0}{s} \right)\right) \\
   &{} - \frac{t_5}{T}\omega^\mu\left(\pi_{\mu\nu}\zeta^\nu - \frac{1}{3s}\zeta_\mu u^\nu\del_\nu s\right)
       - \frac{t^a_6}{T}B_a^\mu\left(\pi_{\mu\nu}\zeta^\nu - \frac{1}{3s}\zeta_\mu u^\nu\del_\nu s\right) \\
   &{} - \frac{t_7}{T}\epsilon^{\mu\nu\rho\sigma}\zeta^\lambda\pi_{\lambda\mu}u_\nu\zeta_\rho\del_\sigma T
       - \frac{t^a_8}{T}\epsilon^{\mu\nu\rho\sigma}\zeta^\lambda\pi_{\lambda\mu}u_\nu\zeta_\rho\nabla_\sigma\frac{\mu_a}{T}
       - \frac{t_9}{T}\epsilon^{\mu\nu\rho\sigma}\zeta^\lambda\pi_{\lambda\mu}u_\nu\zeta_\rho\del_\sigma\zeta^2 \\
   &{} - \frac{t^a_{10}}{T}\epsilon^{\mu\nu\rho\sigma}\zeta^\lambda\pi_{\lambda\mu}u_\nu\zeta_\rho\hat E_{a\sigma}
       - \frac{t_{13}}{T}\epsilon^{\mu\nu\rho\sigma}\pi_\mu^\lambda\nabla_{(\lambda}\xi_{\nu)}u_\rho\zeta_\sigma \\
   &{} + \frac{j^a_1}{T}\omega^\mu\hat E_{a\mu} + \frac{j^{ab}_2}{T}B_b^\mu\hat E_{a\mu}
       + \frac{j^a_3}{T}\zeta_\mu\omega^\mu\zeta^\nu\hat E_{a\nu} + \frac{j^{ab}_4}{T}\zeta_\mu B_b^\mu\zeta^\nu\hat E_{a\nu}
       - \frac{j^a_5}{T}\epsilon^{\mu\nu\rho\sigma}\del_\mu T u_\nu\zeta_\rho\hat E_{a\sigma} \\
   &{} - \frac{j^{ab}_6}{T}\epsilon^{\mu\nu\rho\sigma}\nabla_\mu\frac{\mu_b}{T} u_\nu\zeta_\rho\hat E_{a\sigma}
       - \frac{j^a_7}{T}\epsilon^{\mu\nu\rho\sigma}\del_\mu\zeta^2 u_\nu\zeta_\rho\hat E_{a\sigma}
       - \frac{j^{ab}_8}{T}\epsilon^{\mu\nu\rho\sigma}\hat E_{b\mu} u_\nu\zeta_\rho\hat E_{a\sigma} \\
   &{} - \frac{j^a_9}{T}\epsilon^{\mu\nu\rho\sigma}\zeta^\lambda\pi_{\lambda\mu}u_\nu\zeta_\rho\hat E_{a\sigma} \\
   &{} + \frac{\nu_1 s}{T}\zeta_\mu\omega^\mu u^\nu\del_\nu\frac{n_0}{s} + \frac{\nu^a_2 s}{T}\zeta_\mu B_a^\mu u^\nu\del_\nu\frac{n_0}{s} \\
   &{} + 2s_1\omega^\mu\left(\frac{n_a}{h}\hat E^a_\mu - \frac{1}{T}\del_\mu T - \frac{s}{h}\zeta_\mu u^\nu\del_\nu\frac{n_0}{s}\right) + \omega^\mu\del_\mu s_1 \\
   &{} + s^a_2\left(B_a^\mu\left(\frac{n_b}{h}\hat E^b_\mu - \frac{1}{T}\del_\mu T - \frac{s}{h}\zeta_\mu u^\nu\del_\nu\frac{n_0}{s}\right) - 2E_{a\mu}\omega^\mu\right)
       + B_a^\mu\nabla_\mu s^a_2 \\
   &{} - s^a_3\left(E_\mu^0 B^\mu_a + E_{a\mu} B_0^\mu \right) + \left(\zeta_\nu B_a^\nu u^\mu - \mu_0 B_a^\mu + \epsilon^{\nu\rho\sigma\mu}E_{a\nu}u_\rho\zeta_\sigma \right)
          \nabla_\mu s^a_3 \\
   &{} + s_4\del_\mu T\left(2\mu_0\omega^\mu + B_0^\mu - 2u^\mu\zeta_\nu\omega^\nu - \frac{n_a}{h}\epsilon^{\mu\nu\rho\sigma}u_\nu\zeta_\rho\hat E^a_\sigma \right)
         + \epsilon^{\mu\nu\rho\sigma}\del_\mu s_4 u_\nu\zeta_\rho\del_\sigma T \\
   &{} + s^a_5\nabla_\mu\frac{\mu_a}{T}\left(2\mu_0\omega^\mu + B_0^\mu - 2u^\mu\zeta_\nu\omega^\nu
         - \epsilon^{\mu\nu\rho\sigma}u_\nu\zeta_\rho\left(\frac{n_b}{h}\hat E^b_\sigma - \frac{1}{T}\del_\sigma T\right)\right)
         + \epsilon^{\mu\nu\rho\sigma}\nabla_\mu s^a_5 u_\nu\zeta_\rho\nabla_\mu\frac{\mu_a}{T} \\
   &{} + s_6\del_\mu\zeta^2\left(2\mu_0\omega^\mu + B_0^\mu - 2u^\mu\zeta_\nu\omega^\nu
         - \epsilon^{\mu\nu\rho\sigma}u_\nu\zeta_\rho\left(\frac{n_a}{h}\hat E^a_\sigma - \frac{1}{T}\del_\sigma T \right)\right)
         + \epsilon^{\mu\nu\rho\sigma}\del_\mu s_6 u_\nu\zeta_\rho\del_\sigma\zeta^2 \\
   &{} + s_7\left(\zeta_\mu\omega^\mu\left(s\left(\frac{1}{Q} + \frac{\mu_0^2}{h}\right) u^\nu\del_\nu\frac{n_0}{s} - \frac{\mu_0}{3s}u^\nu\del_\nu s
         - \frac{1}{Q}\xi^\nu\del_\nu Q\right) \right. \\
   &\quad\left.{} + \omega^\mu\left(\mu_0\hat E^0_\mu - \frac{\mu_0^2 n_a}{h}\hat E^a_\mu + \frac{1}{2}\del_\mu\zeta^2 \right)
         + \zeta^\mu\omega^\nu(B^0_{\mu\nu} + \mu_0\pi_{\mu\nu}) \right. \\
   &\quad\left.{} + \frac{1}{2}\epsilon^{\mu\nu\rho\sigma}\zeta_\mu u_\nu\left(\zeta^\lambda(\pi_{\lambda\rho} + \omega_{\lambda\rho})
         \left(\frac{n_a}{h}\hat E^a_\sigma - \frac{1}{T}\del_\sigma T \right) + \xi^\lambda\nabla_\lambda\nabla_\rho u_\sigma \right)\right)
         + \zeta_\mu\omega^\mu\xi^\nu\del_\nu s_7 \\
   &{} + s^a_8\left(\zeta_\mu B_a^\mu\left(s\left(\frac{1}{Q} + \frac{\mu_0^2}{h}\right) u^\nu\del_\nu\frac{n_0}{s} - \frac{\mu_0}{3s}u^\nu\del_\nu s
         - \frac{1}{Q}\xi^\nu\del_\nu Q\right) \right. \\
   &\quad\left.{} + B_a^\mu\left(\mu_0\hat E^0_\mu - \frac{\mu_0^2 n_b}{h}\hat E^b_\mu + \frac{1}{2}\del_\mu\zeta^2 \right)
         + \zeta^\mu B_a^\nu\left(B^0_{\mu\nu} + \mu_0(\pi_{\mu\nu} + \omega_{\mu\nu})\right) \right. \\
   &\quad\left.{} + \epsilon^{\mu\nu\rho\sigma}\zeta_\mu u_\nu E_{a\rho}\left(\mu_0\left(\frac{1}{T}\del_\sigma T - \frac{n_b}{h}\hat E^b_\sigma\right)
         + \zeta^\lambda(\pi_{\lambda\sigma} + \omega_{\lambda\sigma})\right)
       + \frac{1}{2}\epsilon^{\mu\nu\rho\sigma}\zeta_\mu u_\nu\xi^\lambda\nabla_\lambda F_{a\rho\sigma} \right) \\
   &\quad{} + \zeta_\mu B_a^\mu\xi^\nu\del_\nu s^a_8 \\
   &{} + s^a_9\left(\epsilon^{\mu\nu\rho\sigma}\zeta_\mu u_\nu\left(E_{a\rho}\left(\frac{n_b}{h}\hat E^b_\sigma - \frac{1}{T}\del_\sigma T \right)
         + (\pi_\rho{}^\lambda + \omega_\rho{}^\lambda)B_{a\lambda\sigma} + \frac{1}{2}u^\lambda\nabla_\lambda F_{a\rho\sigma} \right) \right. \\
   &\quad\left.{} + E_{a\mu}(2\mu_0\omega^\mu + B_0^\mu) - \frac{2}{3s}\zeta_\mu B_a^\mu u^\nu\del_\nu s \right)
       + \epsilon^{\mu\nu\rho\sigma}\nabla_\mu s^a_9 u_\nu\zeta_\rho E_{a\sigma} \\
   &{} + s_{10}\left(2\zeta_\mu\omega^\mu\left(\zeta^\nu\left(\frac{1}{T}\del_\nu T - \frac{n_a}{h}\hat E^a_\nu\right) + \frac{s\zeta^2}{h}u^\nu\del_\nu\frac{n_0}{s}
         - \frac{\mu_0}{3s}u^\nu\del_\nu s \right) + 2\mu_0\omega^\mu\zeta^\nu\pi_{\mu\nu} \right. \\
   &\quad\left.{} + B_0^\mu\zeta^\nu(\pi_{\mu\nu} + \omega_{\mu\nu})
        + \epsilon^{\mu\nu\rho\sigma}\zeta_\mu u_\nu(\pi_\rho{}^\lambda + \omega_\rho{}^\lambda)
           \left(\zeta_\lambda\left(\frac{n_a}{h}\hat E^a_\sigma - `\frac{1}{T}\del_\sigma T \right)
         + \nabla_{(\sigma}\xi_{\lambda)} + \frac{1}{2}B^0_{\sigma\lambda} \right) \right. \\
   &\quad\left.{} - \frac{1}{2}\epsilon^{\mu\nu\rho\sigma}\zeta_\mu u_\nu R_{\rho\sigma\lambda\kappa}\zeta^\lambda u^\kappa \right)
         + \epsilon^{\mu\nu\rho\sigma}\del_\mu s_{10}u_\nu\zeta_\rho(\pi_{\sigma\lambda} + \omega_{\sigma\lambda})\zeta^\lambda \ .
\end{align*}

\subsection{Cancellation requirements} \label{app:calc:cancellation}

The $s_7$, $s^a_8$ and $s^a_9$ terms contain distinct second-derivative terms. Therefore, $s_7 = s_8^a = s_9^a = 0$. The Riemann-curvature contribution from the $s_{10}$ term cannot be canceled with anything, so we have $s_{10} = 0$.

$t_5$ and $t^a_6$ vanish as the only terms containing $\pi_{\mu\nu}\zeta^\mu\omega^\nu$ and $\pi_{\mu\nu}\zeta^\mu B_a^\nu$ respectively. Similarly, the $t_7$, $t_8^a$, $t_9$ and $t_{13}$ terms cannot be canceled with anything, so they vanish. $t_{11} \equiv a$ and $t_{12} \equiv \chi$ don't contribute at all to the entropy production rate. They are therefore arbitrary functions of state. If we consider only the parity-odd sector, then the contribution from the $t^a_{10}$ term must cancel with the corresponding contribution from the $j^a_9$ term, giving $t^a_{10} = -j^a_9$. However, it was noticed in \cite{Bhattacharya:2011tr} that these terms can be counterbalanced by contributions from the parity-even sector. We then have two arbitrary functions $t^a_{10} \equiv b^a_1$ and $j^a_9 \equiv b^a_2$. Given time-reversal invariance, the Onsager principle actually gives $b_2^a = -b_1^a$, removing the need for counterbalancing from the parity-even sector; see section \ref{sec:results:onsager}.

We now turn to the contributions to $\nabla_\mu s^\mu$ without any factors of $\zeta_\mu$. Collecting the coefficients of $\omega^\mu\del_\mu T$, $\omega^\mu\nabla_\mu(\mu^a/T)$, $\omega^\mu\del_\mu\zeta^2$, $\omega^\mu\hat E^a_\mu$, $B_a^\mu\del_\mu T$, $B_a^\mu\nabla_\mu(\mu^b/T)$, $B_a^\mu\del_\mu\zeta^2$ and $B_a^\mu\hat E^b_\mu$, we get a generalization of the differential equations in \cite{Son:2009tf}:
\begin{align}
 \begin{split}
   & \dd{s_1}{T}{\mu_a/T,\zeta^2} - \frac{2s_1}{T} + 2\mu_0 s_4 = 0; \\
   & \dd{s_1}{(\mu_a/T)}{T,\zeta^2} - 2Ts^a_2 + 2\mu_0 s^a_5 = 0; \\
   & \dd{s_1}{\zeta^2}{T,\mu_a/T} + 2\mu_0 s_6 = 0; \\
   & \frac{2n^a s_1}{h} - 2s^a_2 + \frac{j^a_1}{T} = 0; \\
   & \dd{s^a_2}{T}{\mu_b/T,\zeta^2} - \frac{s^a_2}{T} + \delta^a_0 s_4 - \mu_0\dd{s^a_3}{T}{\mu_b/T,\zeta^2} = 0; \\
   & \dd{s^a_2}{(\mu_b/T)}{T,\zeta^2} + \delta^a_0 s^b_5 - 2T\delta_0^{(a}s_3^{b)} - \mu_0\dd{s^a_3}{(\mu_b/T)}{T,\zeta^2} - C^{abc}\mu_c = 0; \\
   & \dd{s^a_2}{\zeta^2}{T,\mu_b/T} + \delta^a_0 s_6 - \mu_0\dd{s^a_3}{\zeta^2}{T,\mu_b/T} = 0; \\
   & \frac{n^a s^b_2}{h} + \frac{j^{ab}_2}{T} - C^{abc}\frac{\mu_c}{T} - 2\delta_0^{(a}s_3^{b)} = 0\ ,
 \end{split}
\end{align}
which can be rewritten as:
\begin{align}
   & \dd{(s_1/T^2)}{T}{\mu_a/T,\zeta^2} + \frac{2\mu_0 s_4}{T^2} = 0; \label{eq:ds1_dT} \\
   & \dd{(s_1/T^2)}{(\mu_a/T)}{T,\zeta^2} + \frac{2\mu_0 s^a_5}{T^2} = \frac{2s^a_2}{T}; \label{eq:ds1_dmu} \\
   & \dd{(s_1/T^2)}{\zeta^2}{T,\mu_a/T} + \frac{2\mu_0 s_6}{T^2} = 0; \label{eq:ds1_dzeta} \\
   & j^a_1 = 2T\left(s^a_2 - \frac{n^a}{h} s_1\right); \label{eq:j1} \\
   & \dd{(s^a_2/T)}{T}{\mu_b/T,\zeta^2} + \delta^a_0 \frac{s_4}{T} - \frac{\mu_0}{T}\dd{\alpha^a}{T}{\mu_b/T,\zeta^2} = 0; \label{eq:ds2_dT} \\
   & \dd{(s^a_2/T)}{(\mu_b/T)}{T,\zeta^2} + \delta^a_0 \frac{s^b_5}{T} - \frac{\mu_0}{T}\dd{\alpha^a}{(\mu_b/T)}{T,\zeta^2} = C^{abc}\frac{\mu_c}{T} + 2\delta_0^{(a}\alpha^{b)}; \label{eq:ds2_dmu} \\
   & \dd{(s^a_2/T)}{\zeta^2}{T,\mu_b/T} + \delta^a_0 \frac{s_6}{T} - \frac{\mu_0}{T}\dd{\alpha^a}{\zeta^2}{T,\mu_b/T} = 0; \label{eq:ds2_dzeta} \\
   & j^{ab}_2 = C^{abc}\mu_c + 2T\delta_0^{(a}\alpha^{b)} - \frac{n^a}{h}Ts^b_2\ . \label{eq:j2}
\end{align}
We renamed $s_3^a \equiv \alpha^a$, since it will turn out that $s_3^a$ is an unconstrained function of state. Eqs. \eqref{eq:ds2_dT}-\eqref{eq:ds2_dzeta} imply that:
\begin{align}
 d\frac{s^a_2}{T} + \frac{1}{T}\delta^a_0\left(s_4 dT + s^b_5 d\frac{\mu_b}{T} + s_6 d\zeta^2 \right) = C^{abc}\frac{\mu_b}{T}d\frac{\mu_c}{T} + \frac{\mu_0}{T}d\alpha^a
   + 2\delta_0^{(a}\alpha^{b)} d\frac{\mu_b}{T}\ .
\end{align}
From this we get:
\begin{align}
 & \frac{s^a_2}{T} = \frac{1}{2T^2}C^{abc}\mu_b\mu_c + 2\delta^{(a}_0\alpha^{b)}\frac{\mu_b}{T} + \beta^a \\
 & s_4 dT + s^a_5 d\frac{\mu_a}{T} + s_6 d\zeta^2 = -T\left(\frac{\mu_a}{T}d\alpha^a + d\beta_0 \right)\ , \label{eq:beta}
\end{align}
where $\beta_0$ is an arbitrary function of state and $\beta_i$ are arbitrary constants. Similarly, eqs. \eqref{eq:ds1_dT}-\eqref{eq:ds1_dzeta} now become:
\begin{align}
 d\frac{s_1}{T^2} + 2\frac{\mu_0}{T^2}\left(s_4 dT + s^a_5 d\frac{\mu_a}{T} + s_6 d\zeta^2\right) = \frac{2s^a_2}{T}d\frac{\mu_a}{T}\ ,
\end{align}
which can be rewritten as:
\begin{align}
 & d\frac{s_1}{T^2} - 2\frac{\mu_0}{T}\left(\frac{\mu_a}{T}d\alpha^a + d\beta_0\right)
    = \left(\frac{1}{T^2}C^{abc}\mu_b\mu_c + 4\delta^{(a}_0\alpha^{b)}\frac{\mu_b}{T} + 2\beta^a\right)d\frac{\mu_a}{T} \\
 & d\frac{s_1}{T^2} = d\left(\frac{1}{3T^3}C^{abc}\mu_a\mu_b\mu_c + \frac{2}{T^2}\alpha^a\mu_a\mu_0 + \frac{2}{T}\beta^a\mu_a\right) \\
 & \frac{s_1}{T^2} = \frac{1}{3T^3}C^{abc}\mu_a\mu_b\mu_c + \frac{2}{T^2}\alpha^a\mu_a\mu_0 + \frac{2}{T}\beta^a\mu_a + \gamma\ ,
\end{align}
where $\gamma$ is another arbitrary integration constant. As noticed in \cite{Bhattacharya:2011tr}, CPT invariance requires $\gamma = 0$. Summing up and using eqs. \eqref{eq:j1} and \eqref{eq:j2} for $j^a_1$ and $j^{ab}_2$, we have:
\begin{align}
 s_1 ={}& \frac{1}{3T}C^{abc}\mu_a\mu_b\mu_c + 2\alpha^a\mu_a\mu_0 + 2\beta^a\mu_a T + \gamma T^2 \\
 s^a_2 ={}& \frac{1}{2T}C^{abc}\mu_b\mu_c + 2\delta^{(a}_0\alpha^{b)}\mu_b + \beta^a T \\
 \begin{split}
   j^a_1 ={}& C^{abc}\mu_b\mu_c + 4\delta^{(a}_0\alpha^{b)}\mu_b T + 2\beta^a T^2 \\
      &- \frac{2n^a}{h}\left(\frac{1}{3}C^{bcd}\mu_b\mu_c\mu_d + 2\alpha^b\mu_b\mu_0 T + 2\beta^b\mu_b T^2 + \gamma T^3\right)
 \end{split} \\
 \begin{split}
   j^{ab}_2 ={}& C^{abc}\mu_c + 2T\delta_0^{(a}\alpha^{b)} - \frac{n^a}{h}\left(\frac{1}{2}C^{bcd}\mu_c\mu_d + 2\delta^{(b}_0\alpha^{c)}\mu_c T + \beta^b T^2 \right)\ .
 \end{split}
\end{align}

We now collect the contributions to $\nabla_\mu s^\mu$ of the form $\epsilon^{\mu\nu\rho\sigma}U_\mu u_\nu\xi_\rho V_\sigma$, where $U_\mu$ and $V_\mu$ are combinations of the independent vectors $\del_\mu p$, $\del_\mu\xi$, $\nabla_\mu(\mu_a/T)$ and $\hat E^a_\mu$. For brevity, we will use the wedge notation $\mathbf{U}\wedge\mathbf{V}$, and write gradients as exterior derivatives. On charged quantities, we will use the gauge-covariant exterior derivative $D$, with e.g. $D^2\mu_a = f_{abc}F^b\mu^c$. The cancellation requirement on these terms reads:
\begin{align}
 \begin{split}
   & \frac{j_8^{ab}}{T} \mathbf{\hat E}_a\wedge\mathbf{\hat E}_b + \left(\frac{j^a_5}{T} + \frac{n^a s_4}{h} + \dd{\alpha^a}{T}{\mu_b/T,\zeta^2}\right)\mathbf{\hat E}_a\wedge dT \\
   &{} + \left(\frac{j^{ab}_6}{T} + \frac{n^a s^b_5}{h} + \dd{\alpha^a}{(\mu_b/T)}{T,\zeta^2}\right)\mathbf{\hat E}_a\wedge D\frac{\mu_b}{T}
    + \left(\frac{j^a_7}{T} + \frac{n^a s_6}{h} + \dd{\alpha^a}{\zeta^2}{T,\mu_b/T}\right)\mathbf{\hat E}_a\wedge d\zeta^2 \\
   &{} - \frac{s^a_5}{T} dT\wedge D\frac{\mu_a}{T} - \frac{s_6}{T} dT\wedge d\zeta^2 + ds_4\wedge dT + Ds^a_5\wedge D\frac{\mu_a}{T} + ds_6\wedge d\zeta^2
    + T D\frac{\mu_a}{T}\wedge D\alpha^a = 0\ . \label{eq:wedges}
 \end{split}
\end{align}
Considering entropy contributions from the parity-odd sector alone, the first term in \eqref{eq:wedges} must vanish, giving $j_8^{[ab]} = 0$, with $j_8^{(ab)}$ unconstrained. However, due to possible mixing with the parity-even sector, all of $j_8^{ab} \equiv c^{ab}$ can be arbitrary; this is a straightforward generalization of the situation with the $\epsilon^{\mu\nu\rho\sigma}\hat E^a_\mu u_\nu\zeta_\rho\pi_{\sigma\lambda}\zeta^\lambda$ contributions in \cite{Bhattacharya:2011tr}. On the other hand, given time-reversal invariance, $c_{ab}$ is in fact symmetric due to the Onsager principle, making the mixing with the parity-even sector irrelevant. The other terms in the first two lines of \eqref{eq:wedges} must vanish separately, giving:
\begin{align}
 \begin{split}
   & j^a_5 = -T\left(\frac{n^a s_4}{h} + \dd{\alpha^a}{T}{\mu_b/T,\zeta^2}\right);\quad j^{ab}_6 = -T\left(\frac{n^a s^b_5}{h} + \dd{\alpha^a}{(\mu_b/T)}{T,\zeta^2}\right); \\
   & j^a_7 = -T\left(\frac{n^a s_6}{h} + \dd{\alpha^a}{\zeta^2}{T,\mu_b/T}\right)\ .
 \end{split}
\end{align}
Using eq. \eqref{eq:beta}, we write the vanishing of the last line of \eqref{eq:wedges} as:
\begin{align}
 \begin{split}
   0 &= dT\wedge\left(\frac{\mu_a}{T}D\alpha^a + d\beta_0\right) + d\left(s_4 dT + s_5^a D\frac{\mu_a}{T} + s_6 d\zeta^2 \right) - s_5^a D^2\frac{\mu_a}{T}
       + T D\frac{\mu_a}{T}\wedge D\alpha^a \\
     &= dT\wedge\left(\frac{\mu_a}{T}D\alpha^a + d\beta_0\right) - d\left(T\left(\frac{\mu_a}{T}D\alpha^a + d\beta_0 \right)\right) - s_5^a D^2\frac{\mu_a}{T}
       + T D\frac{\mu_a}{T}\wedge D\alpha^a \\
     &= -\mu_a D^2\alpha^a - s_5^a D^2\frac{\mu_a}{T} = -\mu^a f_{abc} F^b\alpha^c -\frac{1}{T} s_5^a f_{abc}F^b\mu^c\ .
 \end{split} \label{eq:identity}
\end{align}
The $f_{abc}$ terms vanish, because $\alpha^a$ and $s_5^a$ are functions of $\mu_a$ plus scalars, and therefore commute with $\mu_a$ under the charge group. Thus, \eqref{eq:identity} is an identity, and doesn't impose any further constraints.

We proceed to examine the remaining contributions to $\nabla_\mu s^\mu$. Cancellation of the $\zeta_\mu\omega^\mu\zeta^\nu\hat E^a_\nu$ factors requires $j^a_3 = (\mu_0 n^a/h - \delta_0^a)t_3$. Similarly, the $\zeta_\mu B_b^\mu\zeta^\nu\hat E^a_\nu$ factors give $j^{ab}_4 = (\mu_0 n^a/h - \delta_0^a)t^b_4$. Collecting the factors of the form $\zeta_\mu\omega^\mu u^\nu\nabla_\nu x$ in $\nabla_\mu s^\mu$, we get:
\begin{align}
 \begin{split}
   0 ={}& u^\mu\left(\frac{t_1}{Ts}\del_\mu s + \frac{t_3}{2T}\del_\mu\zeta^2
      + \frac{s}{Th}\left(\mu_0\zeta^2 t_3 + h\nu_1 - 2Ts_1\right)\del_\mu\frac{n_0}{s} \right. \\
   &\left.{}\vphantom{\frac{t_1}{T}} - 2s_4\del_\mu T - 2s^a_5\nabla_\mu\frac{\mu_a}{T} - 2s_6\del_\mu\xi^2 \right) \\
     ={}&  u^\mu\left(\frac{t_1}{Ts}\del_\mu s + \frac{t_3}{2T}\del_\mu\zeta^2 + \frac{s}{Th}\left(\mu_0\zeta^2 t_3 + h\nu_1 - 2Ts_1\right)\del_\mu\frac{n_0}{s}
      + 2T\left(\frac{\mu_a}{T}\nabla_\mu\alpha^a + \del_\mu\beta_0 \right)\right)\ .
 \end{split}
\end{align}
Recalling that $u^\mu\nabla_\mu(n_i/s)$ vanishes due to the ideal equations, we can write this condition as:
\begin{align}
 0 = \frac{t_1}{Ts}ds + \frac{t_3}{2T}d\zeta^2 + \frac{s}{Th}\left(\mu_0\zeta^2 t_3 + h\nu_1 - 2Ts_1\right)d\frac{n_0}{s} + \lambda_i D\frac{n_i}{s}
    + 2T\left(\frac{\mu_a}{T}D\alpha^a + d\beta_0 \right)\ ,
\end{align}
with arbitrary functions $\lambda_i$. Choosing $(s, n_a/s, \zeta^2)$ as the independent thermal parameters (and discarding the previous basis $(T, \mu_a/T, \zeta^2)$), we get:
\begin{align}
 t_1 ={}& {-2sT^2} \left(\frac{\mu_a}{T}\ddsimple{\alpha^a}{s} + \ddsimple{\beta_0}{s}\right) \\
 t_3 ={}& {-4T^2} \left(\frac{\mu_a}{T}\ddsimple{\alpha^a}{\zeta^2} + \ddsimple{\beta_0}{\zeta^2} \right) \\
 \begin{split}
   \nu_1 ={}& \frac{2T}{h}s_1 - \frac{\mu_0\zeta^2}{h}t_3 - \frac{2T^2}{s}\left(\frac{\mu_a}{T}\ddsimple{\alpha^a}{(n_0/s)} + \ddsimple{\beta_0}{(n_0/s)}\right) \\
     ={}& \frac{2}{h}\left(\frac{1}{3}C^{abc}\mu_a\mu_b\mu_c + 2\alpha^a\mu_a\mu_0 T + 2\beta^a\mu_a T^2 + \gamma T^3 \right) \\
      &+ \frac{4T^2\mu_0\zeta^2}{h}\left(\frac{\mu_a}{T}\ddsimple{\alpha^a}{\zeta^2} + \ddsimple{\beta_0}{\zeta^2} \right)
      - \frac{2T^2}{s}\left(\frac{\mu_a}{T}\ddsimple{\alpha^a}{(n_0/s)} + \ddsimple{\beta_0}{(n_0/s)}\right)\ .
 \end{split}
\end{align}
Similarly, for factors of the form $\xi_\mu B_a^\mu u^\nu\nabla_\nu x$, we have:
\begin{align}
 0 = \frac{t^a_2}{Ts}ds + \frac{t^a_4}{2T}d\zeta^2 + \frac{s}{Th}\left(\mu_0\zeta^2 t^a_4 + h\nu^a_2 - Ts^a_2 \right)d\frac{n_0}{s}
    + \lambda^a_i D\frac{n_i}{s} + D\alpha^a\ ,
\end{align}
with arbitrary $\lambda^a_i$. Again using $(s, n_a/s, \zeta^2)$ as a set of independent thermal parameters, we get:
\begin{align}
 t^a_2 &= -sT\ddsimple{\alpha^a}{s} \\
 t^a_4 &= -2T\ddsimple{\alpha^a}{\zeta^2} \\
 \begin{split}
   \nu^a_2 &=  \frac{T}{h}s^a_2 - \frac{\mu_0\zeta^2}{h}t_4^a - \frac{T}{s}\ddsimple{\alpha^a}{(n_0/s)} \\
     &= \frac{1}{h}\left(\frac{1}{2}C^{abc}\mu_b\mu_c + 2\delta_0^{(a}\alpha^{b)}\mu_b T + \beta^a T^2\right) + \frac{2T\mu_0\zeta^2}{h}\ddsimple{\alpha^a}{\zeta^2}
      - \frac{T}{s}\ddsimple{\alpha^a}{(n_0/s)}\ .
 \end{split}
\end{align}
This concludes the analysis of all the terms in the entropy production rate. Putting everything together, we arrive at the result \eqref{eq:T_1}-\eqref{eq:s_1}.


\begin{thebibliography} {99}

\bibitem{Kapitza} P. Kapitza, "Viscosity of Liquid Helium below the $\lambda$-Point",
 Nature {\bf 141} (1938) 74.

\bibitem{Landau} L. D. Landau, "The Theory of Superfluidity of Helium II", J. Phys. USSR {\bf 5} (1941) 75.

\bibitem{Tisza} L. Tisza, "Transport Phenomena in Helium II", Nature {\bf 141} (1938) 913.

\bibitem{relativistic}
  D.~T.~Son,
  ``Hydrodynamics of relativistic systems with broken continuous symmetries,''
  Int.\ J.\ Mod.\ Phys.\  A {\bf 16S1C}, 1284 (2001)
  [arXiv:hep-ph/0011246];
  C.~Pujol and D.~Davesne,
  ``Relativistic dissipative hydrodynamics with spontaneous symmetry breaking,''
  Phys.\ Rev.\  C {\bf 67}, 014901 (2003)
  [arXiv:hep-ph/0204355];
  M.~A.~Valle,
  ``Hydrodynamic fluctuations in relativistic superfluids,''
  Phys.\ Rev.\  D {\bf 77}, 025004 (2008)
  [arXiv:0707.2665 [hep-ph]].

\bibitem{NStars}
  M.~E.~Gusakov and N.~Andersson,
  ``Temperature dependent pulsations of superfluid neutron stars,''
  Mon.\ Not.\ Roy.\ Astron.\ Soc.\  {\bf 372}, 1776 (2006)
  [arXiv:astro-ph/0602282];
  M.~E.~Gusakov,
  ``Bulk viscosity of superfluid neutron stars,''
  Phys.\ Rev.\  D {\bf 76}, 083001 (2007)
  [arXiv:0704.1071 [astro-ph]].

\bibitem{CFL}
  M.~Mannarelli and C.~Manuel,
  ``Bulk viscosities of a cold relativistic superfluid: Color-flavor locked quark matter,''
  Phys.\ Rev.\  D {\bf 81}, 043002 (2010)
  [arXiv:0909.4486 [hep-ph]].

\bibitem{Alford:1997zt}
  M.~G.~Alford, K.~Rajagopal, F.~Wilczek,
  ``QCD at finite baryon density: Nucleon droplets and color superconductivity,''
  Phys.\ Lett.\  {\bf B422}, 247-256 (1998).
  [hep-ph/9711395].

\bibitem{Alford:2007xm}
  M.~G.~Alford, A.~Schmitt, K.~Rajagopal and T.~Schafer,
  ``Color superconductivity in dense quark matter,''
  Rev.\ Mod.\ Phys.\  {\bf 80}, 1455 (2008)
  [arXiv:0709.4635 [hep-ph]].

\bibitem{Bhattacharya:2011ee}
  J.~Bhattacharya, S.~Bhattacharyya and S.~Minwalla,
  ``Dissipative Superfluid dynamics from gravity,''
  arXiv:1101.3332 [hep-th].

\bibitem{Herzog:2011ec}
  C.~P.~Herzog, N.~Lisker, P.~Surowka and A.~Yarom,
  ``Transport in holographic superfluids,''
  arXiv:1101.3330 [hep-th].

\bibitem{Son:2009tf}
  D.~T.~Son and P.~Surowka,
  ``Hydrodynamics with Triangle Anomalies,''
  Phys.\ Rev.\ Lett.\  {\bf 103}, 191601 (2009)
  [arXiv:0906.5044 [hep-th]].

\bibitem{Neiman:2010zi}
  Y.~Neiman and Y.~Oz,
  ``Relativistic Hydrodynamics with General Anomalous Charges,''
  JHEP {\bf 1103} (2011) 023
  [arXiv:1011.5107 [hep-th]].

\bibitem{Loganayagam:2011mu}
  R.~Loganayagam,
  ``Anomaly Induced Transport in Arbitrary Dimensions,''
  arXiv:1106.0277 [hep-th].

\bibitem{KharzeevSecondOrder:2011ds}
  D.~E.~Kharzeev and H.~U.~Yee,
  ``Anomalies and time reversal invariance in relativistic hydrodynamics: the second order and higher dimensional formulations,''
  arXiv:1105.6360 [hep-th].

\bibitem{Sadofyev:2010is}
  A.~V.~Sadofyev, V.~I.~Shevchenko and V.~I.~Zakharov,
  ``Notes on chiral hydrodynamics within effective theory approach,''
  Phys.\ Rev.\  D {\bf 83}, 105025 (2011)
  [arXiv:1012.1958 [hep-th]].

\bibitem{Erdmenger:2008rm}
  J.~Erdmenger, M.~Haack, M.~Kaminski and A.~Yarom,
  ``Fluid dynamics of R-charged black holes,''
  JHEP {\bf 0901}, 055 (2009)
  [arXiv:0809.2488 [hep-th]].

\bibitem{Banerjee:2008th}
  N.~Banerjee, J.~Bhattacharya, S.~Bhattacharyya, S.~Dutta, R.~Loganayagam and P.~Surowka,
  ``Hydrodynamics from charged black branes,''
  arXiv:0809.2596 [hep-th].

\bibitem{Eling:2010hu}
  C.~Eling, Y.~Neiman and Y.~Oz,
  ``Holographic Non-Abelian Charged Hydrodynamics from the Dynamics of Null Horizons,''
  JHEP {\bf 1012} (2010) 086
  [arXiv:1010.1290 [hep-th]].

\bibitem{KerenZur:2010zw}
  B.~Keren-Zur, Y.~Oz,
  ``Hydrodynamics and the Detection of the QCD Axial Anomaly in Heavy Ion Collisions,''
  JHEP {\bf 1006}, 006 (2010).
  [arXiv:1002.0804 [hep-ph]].

\bibitem{Kharzeev:2010gr}
  D.~E.~Kharzeev and D.~T.~Son,
  ``Testing the chiral magnetic and chiral vortical effects in heavy ion collisions,''
  arXiv:1010.0038 [hep-ph].

\bibitem{Bhattacharya:2011tr}
  J.~Bhattacharya, S.~Bhattacharyya, S.~Minwalla and A.~Yarom,
  ``A theory of first order dissipative superfluid dynamics,''
  arXiv:1105.3733 [hep-th].

\bibitem{Lin:2011mr}
  S.~Lin,
  ``On the anomalous superfluid hydrodynamics,''
  arXiv:1104.5245 [hep-ph].

\bibitem{Amado:2011zx}
  I.~Amado, K.~Landsteiner, F.~Pena-Benitez,
  ``Anomalous transport coefficients from Kubo formulas in Holography,''
  JHEP {\bf 1105}, 081 (2011).
  [arXiv:1102.4577 [hep-th]].

\bibitem{Landsteiner:2011cp}
  K.~Landsteiner, E.~Megias and F.~Pena-Benitez,
  ``Gravitational Anomaly and Transport,''
  arXiv:1103.5006 [hep-ph].

\bibitem{Landsteiner:2011iq}
  K.~Landsteiner, E.~Megias, L.~Melgar and F.~Pena-Benitez,
  ``Holographic Gravitational Anomaly and Chiral Vortical Effect,''
  arXiv:1107.0368 [hep-th].

\bibitem{onsager} L. Onsager,
"Reciprocal Relations in Irreversible Processes. I.", 
Phys. Rev. {\bf 37}, 405 (1931).

\end{thebibliography}
\end{document}